\documentstyle[sprocl,epsf]{article}
\newcommand{\be}{\begin{equation}}
\newcommand{\ee}{\end{equation}}
\newcommand{\bdm}{\begin{displaymath}}
\newcommand{\edm}{\end{displaymath}}
\newcommand{\bea}{\begin{eqnarray}}
\newcommand{\eea}{\end{eqnarray}}
\newcommand{\no}{\nonumber \\}
\newcommand{\fs}{\; \; .}
\newcommand{\co}{\; \; ,}
\newcommand{\al}{&\!\!\!\!}
\newcommand{\eff}{{e\hspace{-0.1em}f\hspace{-0.18em}f}}

\newcommand{\ren}{{r\hspace{-0.05em}e\hspace{-0.03em}n}}
\newcommand{\ind}{\scriptscriptstyle}
\newcommand{\QCD}{\mbox{\tiny QCD}}
\newcommand{\indR}{\mbox{\tiny R}}
\newcommand{\indL}{\mbox{\tiny L}}
\newcommand{\indV}{\mbox{\tiny V}}
\newcommand{\indA}{\mbox{\tiny A}}
\newcommand{\lvac}{\langle 0|\,}
\newcommand{\rvac}{\,|0\rangle}
\newcommand{\wave}{\raisebox{0.22em}{\fbox{\rule[0.15em]{0em}{0em}\,}}\,}

\newcommand{\qbar}{\overline{\rule[0.42em]{0.4em}{0em}}\hspace{-0.45em}q}
\newcommand{\ubar}{\overline{\rule[0.42em]{0.4em}{0em}}\hspace{-0.5em}u}
\newcommand{\dbar}{\,\overline{\rule[0.65em]{0.4em}{0em}}\hspace{-0.6em}d}
\newcommand{\sbar}{\overline{\rule[0.42em]{0.4em}{0em}}\hspace{-0.5em}s}

\newcommand{\br}{\langle}
\newcommand{\ke}{\rangle}
\newcommand{\Mo}{{M\hspace{-0.62em}\mbox{\raisebox{0.72em}{\tiny o}}
\hspace{0.3em}}}
\begin{document}

\title{CHIRAL DYNAMICS}
\author{H. LEUTWYLER}
\address{Institute for
Theoretical Physics, University of Bern, Sidlerstr. 5,\\ CH-3012 Bern,
Switzerland}

\maketitle\abstracts{The effective field theory relevant for the analysis of
  QCD at low energies is reviewed. The foundations of the method are discussed
in some detail and a few illustrative examples are described.}

%\tableofcontents
\section{Introduction}
The low energy properties of the strong interactions are governed by a chiral
symmetry. For this reason, the physics of the degrees of freedom
that are relevant in this domain is referred to as {\it chiral dynamics} 
and the corresponding effective field theory is called {\it chiral perturbation
  theory}. Effective field theories play an increasingly important role
in physics, not only in the context of the strong interactions, but also in
other areas: heavy quarks, electroweak symmetry breaking, spin models, 
magnetism,
etc. In fact, one of the remarkable features of effective Lagrangians is their
{\it universality}. A compound like La$_2$CuO$_4$ which develops
two-dimensional antiferromagnetic layers and exhibits superconductivity up to
rather high 
temperatures can be described by an effective field theory that closely
resembles the one relevant for the strong interactions! The reason is that the
effective theory only exploits the symmetry properties of the underlying
theory and does not invoke the dynamics of the fields occurring therein. 
In fact, the development of chiral perturbation theory 
started in the 1960's, at a time when it was rather unclear
whether the strong interactions could at all be described in terms of local
fields. Let me briefly review the history of the subject.

In 1957, Goldberger and Treiman investigated the nucleon matrix element of the
axial current.$\,$\cite{Goldberger Treiman} The exchange of a pion
between the nucleon and the current generates a pole at $t=M_\pi^2$, that is at
a very small value of the momentum transfer. Assuming that the contribution
from this nearby pole dominates the matrix element of the divergence at
$t=0$, they obtained the prediction
\bdm
\label{i1}
   g_{\pi N\bar{N}} = \frac{g_A M_N}{F_\pi} \co
\edm
which determines the strength of the pion-nucleon interaction in terms of
the axial current matrix element $g_A$, the nucleon mass $M_N$ and the pion
decay constant $F_\pi$. Inserting the experimental values available at that
time, they found that the relation is indeed obeyed at the 10 \% level
(in the meantime, the discrepancy has become significantly smaller:
The current experimental information$\,$\cite{MENU99} indicates that the 
relation holds to within about 2 or 3\%, but the value of $g_{\pi N\bar{N}}$
is still subject to sizeable uncertainties). 

In 1960, Nambu then showed that the observed smallness of the
pion mass can be explained on the basis of symmetry 
considerations.$\,$\cite{Nambu}
The argument relies on the fact that continuous symmetries may undergo
spontaneous breakdown. If this happens, the spectrum of the theory necessarily
contains massless particles, called Goldstone bosons, after Goldstone who
established the implications of spontaneous symmetry breakdown in 
mathematically precise form.$\,$\cite{Goldstone} In the case of approximate 
symmetries, spontaneous breakdown gives 
rise to particles that are only approximately massless. According to Nambu, the
pions are light because they are the Goldstone bosons of an approximate
symmetry.

The significance of symmetries in particle physics was known since a long time.
Twenty years earlier, Heisenberg had pointed out that the strong interactions
are invariant under the group SU(2) generated by isospin --- if not exactly
then to a high degree of accuracy.$\,$\cite{Heisenberg} The relevance of
{\it approximate symmetries},
however, only emerged in the beginning of the 1960's,
mainly through the work of Gell-Mann.$\,$\cite{Gell-Mann} 
In particular, Gell-Mann and Ne'eman
showed that the observed pattern of mesonic and baryonic
states can be understood if one assumes the strong interactions to be
approximately invariant under a larger group, SU(3). The extended symmetry,
termed the {\it eightfold way}, contains the isospin rotations as a subgroup.
As the extra generators do not commute with the Hamiltonian, the corresponding
currents are not conserved. 
For the Hamiltonian to be approximately symmetric, their divergence  
must, however, be small; accordingly, such currents were
referred to as ``partially conserved".

The symmetry responsible for the smallness of the pion mass is of a different
type. Since the pions carry negative
parity, the relevant generators must change sign under space reflections:
The corresponding partially conserved currents must be axial vectors.
The assumption that the strong interactions admit Partially
Conserved Axial vector Currents was termed the {\it PCAC hypothesis}.

Although the generators of an approximate symmetry group do not commute
with the Hamiltonian, the commutators of the generators among themselves
are fixed by the structure of the group, irrespective of symmetry breaking.
The corresponding currents therefore obey a set of exact
commutation relations --- {\it current algebra}.
In 1965/66 Adler and Weisberger$\,$\cite{Adler Weisberger} 
established the first quantitative consequences
of current algebra and PCAC, and Weinberg showed that the low energy properties
of the amplitude describing the emission of any number of soft pions as
well as the $\pi\pi$ scattering amplitude can unambiguously be
predicted in this framework.$\,$\cite{Weinberg multipi,Weinberg pipi}

The method used to establish these results was based on an analysis
of the Ward identities obeyed by the Green functions of the axial vector
currents and was rather cumbersome. Weinberg, Wess, Zumino, Schwinger, Chang,
Gursey, Lee and others, however, soon
realized that the same results could be derived in a much simpler way, using
{\it effective Lagrangians}. The general framework underlying this technique
was analyzed by Callan, Coleman, Wess and Zumino$\,$\cite{Callan} in 1969. 
At about the same
time, Dashen, Weinstein, Li and Pagels$\,$\cite{Dashen Weinstein Li Pagels} 
started exploring the consequences of the fact that chiral symmetry is only
an approximate symmetry, investigating the departures from the low energy
theorems of current algebra due to symmetry breaking. A concise
formulation in terms of the effective Lagrangian method
was given by Weinberg$\,$\cite{Weinberg 1979} in 1979.
In the meantime, QCD had been discovered, providing a coherent
conceptual framework for an understanding of the strong interactions. In
particular, this theory at once offered a natural explanation for the empirical
fact that the Hamiltonian of the strong interactions exhibits an approximate
symmetry: The Hamiltonian of QCD possesses an exact chiral symmetry
if the quark masses are set equal to zero -- hence an approximate one if 
these masses happen to be small. 

The pioneering work on chiral dynamics concerned the properties
of pion amplitudes on the mass shell. The key observation which gave birth to
this development is that a suitable effective field
theory involving Goldstone fields automatically generates on-shell
amplitudes that obey the low energy theorems of current
algebra and PCAC. The interaction among the Goldstone bosons is described by
an effective Lagrangian that is invariant under global chiral transformations.
The insight
gained thereby not only led to a considerable simplification of current algebra
calculations, but also paved the way to a systematic investigation of the low
energy structure.

The shortcoming of the on-shell analysis is that it does not allow one to 
evaluate current matrix elements such as $F_\pi$ or $g_A$, which -- as
illustrated by the Goldberger-Treiman relation -- play an important role for
the quantitative consequences of the symmetry properties. 
The problem originates in the fact that the on-shell analysis
is based on {\it global}$\,$ symmetry considerations. Global symmetry
provides important
constraints, but does not suffice to determine the low energy structure
beyond leading order. A conclusive framework only results if the
properties of the theory are analyzed off the mass shell: One needs to
consider Green functions and study the
Ward identities which express the symmetries of the underlying theory at
the {\it local}$\,$ level. The
occurrence of anomalies illustrates the problem: Massless QCD is
invariant under global $\mbox{SU}(\mbox{N}_f)_{\indR}
\times \mbox{SU}(\mbox{N}_f)_{\indL}$, but -- for more than two flavours -- the
corresponding effective Lagrangian is not.

\newpage
In 1983, we proposed a method that incorporates the Ward identities
ab initio and allows one to analyze the low energy structure of the Green
functions in a controlled manner.$\,$\cite{GL 1983,GL 1985} 
We worked out a number of applications and estimated 
the effective couplings occurring at first
nonleading order. Most of the more recent work on chiral dynamics
is based on this framework. 

The following presentation of chiral dynamics
only covers a small fraction of the field.
Throughout, I will restrict myself to the mesonic sector of Hilbert 
space. The extension of chiral perturbation theory to the sectors with nonzero
baryon number has recently attracted considerable attention,
for several reasons -- the beautiful experimental results concerning pion
photo- and electroproduction, the fact that the matrix element 
$\langle N|\, \sbar s\, |N\rangle$ can be determined by 
means of $\pi N$ scattering, the Lorentz invariant formulation of the 
effective theory, baryons at large $N_c$, nuclear forces, to name a few. 
For a review of that development,
I refer to the contributions by A.~Manohar and U.~Meissner in this volume.
Even in the mesonic sector, the presentation is far from complete: 
The extension of the 
effective Lagrangian required to incorporate the degrees of freedom of 
the photons and leptons or to analyze nonleptonic weak transitions,  
will not be discussed. The significant progress in our understanding  
of the isospin breaking effects due
to the mass difference between the $u$-- and $d$--quarks -- of relevance, 
for instance, for the analysis of the ratio $\epsilon'/\epsilon$ -- 
will not be covered, either. For a more complete picture of the current state
of chiral dynamics, I refer to the review articles listed in the 
bibliography.$\,$\cite{reviews} 

\section{Massless QCD -- a theoretical paradise}
For reasons that yet remain to be understood, it so happens that the Yukawa 
interaction of the $u$, $d$ and $s$ quarks with the Higgs
field is weak, while the one of the remaining three quark flavours 
is strong. As a first approximation, we may consider the theoretical 
limit where $m_u$, $m_d$ and $m_s$ are set equal to zero, while $m_c$, $m_b$ 
and $m_t$ are sent to infinity. In this limit, QCD is a paradise of a theory:
It does not contain a single dimensionless parameter. If the momenta are
measured in units of the intrinsic scale of the theory, $\Lambda_{\QCD}$,
all of the transition probabilities of physical interest are unambiguously 
determined by the Lagrangian. 

The Hamiltonian of QCD with three massless quark flavours
is characterized by a high degree of symmetry, which originates
in the fact that the interaction between the quarks and gluons is flavour
independent and preserves helicity: It is invariant under independent flavour 
rotations of the right-
and left-handed quark fields. The eight vector currents as well as the 
eight axial currents
\bea V^\mu_a=\qbar \gamma_\mu\mbox{$\frac{1}{2}$}\lambda_a q\co\hspace{2em}
A^\mu_a=\qbar \gamma_\mu\gamma_5\mbox{$\frac{1}{2}$}\lambda_a q\eea
are conserved. The same holds for the singlet vector current $V^\mu_0$,
while the divergence of $A^\mu_0$ contains an anomaly,
\bea\label{eq:omega} \partial_\mu A^\mu_0=
\sqrt{6}\,\omega\co\hspace{2em}\omega=\frac{1}{16\pi^2}\;\mbox{tr}
\hspace{-0.55em}\rule[-0.5em]{0em}{0em}_c \hspace{0.5em}
G_{\mu\nu}\tilde{G}^{\mu\nu}\co\hspace{2em}\lambda_0=\sqrt{\frac{2}{3}}\fs
\eea
The 9 conserved vector charges $Q^{\indV}_0,\,\ldots\,,\,Q^{\indV}_8$ and
the 8 conserved axial charges $Q^{\indA}_1\,\ldots\,,\,Q^{\indA}_8$ generate
the group G = SU(3)$_{\indR}\times$SU(3)$_{\indL}\times$U(1)$_{\indV}$.

As shown by Vafa and Witten,$\,$\cite{Vafa Witten} the ground state of the
theory is necessarily invariant under the subgroup generated by the
vector charges: $Q^{\indV}_a\rvac=0$. For the axial charges, however,
there are two possibilities: 
\begin{description}
\item{a.} $Q^{\indA}_a\rvac=0$. 
The ground state is invariant under chiral
rotations, G is realized as an ordinary Wigner-Weyl symmetry. The spectrum 
consists of degenerate multiplets that transform irreducibly under G and thus
contain degenerate states of opposite parity.
\item{b.} $Q^{\indA}_a\rvac\neq 0$. The ground state is not symmetric with
  respect to chiral rotations, G is realized as a spontaneously broken or
Nambu-Goldstone symmetry.
The spectrum consists of multiplets of the subgroup that leaves the
vacuum invariant, $\mbox{H}=\mbox{SU(3)}_{\indV}\times\mbox{U(1)}_{\indV}$. 
One of these multiplets consists of the Goldstone bosons: Since the eight
states $Q^{\indA}_a\rvac$ carry the same energy and momentum as the ground
state, the spectrum must contain eight massless
particles.
\end{description}
It still remains to be understood why the minimum of the
energy occurs for an asymmetric rather than a symmetric state, so that
alternative b.~is realized in nature, but there is very strong experimental
evidence for this to be the case. Indeed, the main features of the observed
mass pattern are readily understood on this basis:
The pion mass is small compared to the masses of all other hadrons.
This is to be expected if the strong interactions possess an approximate,
spontaneously broken symmetry with the pions as the corresponding Goldstone
bosons. The Lagrangian of QCD does have the relevant approximate chiral 
symmetry, provided the quark masses $m_u$ and $m_d$ are small.
As an immediate consequence, the strong interactions must approximately 
conserve isospin, because the corresponding symmetry breaking parameter, 
$m_u-m_d$, is then also small.  The particle data tables show that
the levels are grouped in multiplets of SU(3). Since the splitting is much 
larger than the one within the isospin multiplets, the symmetry breaking
parameter $m_s-\frac{1}{2}(m_u+m_d)$ must be large compared to $m_u-m_d$.
The observed pattern thus requires $m_s\gg m_d>m_u$. For the pions to be light
and the eightfold way to be an approximate symmetry, $m_u,\,m_d$ as well
as $m_s-\frac{1}{2}(m_u+m_d)$ must be small. Hence the mass of the strange 
quark must be small, too -- we must be living in a world that is close to
the paradise described above.  

In the real world, chiral symmetry is broken not only
spontaneously, but also explicitly, through the quark mass term in the
Hamiltonian,
\bea H_{\QCD}=H_0+H_1\co\hspace{2em}H_1=\int\!\!d^3\!x\; m_u \ubar u +
m_d \dbar d+ m_s \sbar s\fs\eea
Also, since the heavy quarks are not infinitely heavy, their degrees of
freedom must be included in the Hamiltonian. As these are singlets under the 
group SU(3)$_{\indR}\times$SU(3)$_{\indL}\times$U(1)$_{\indV}$,
they may be booked in $H_0$ -- what counts for the
low energy analysis of the theory is only that $H_0$ is invariant under
this group.

QCD neatly explains why the pseudoscalar octet 
contains the eight lightest
hadrons and why the mass pattern of this multiplet very strongly breaks 
eightfold way symmetry. These particles carry the quantum numbers of the 
Goldstone bosons required by the
spontaneous breakdown of the approximate symmetry
SU(3)$_{\indR}\times$SU(3)$_{\indL}\times$U(1)$_{\indV}
\rightarrow$ SU(3)$_{\indV}\times$U(1)$_{\indV}$. 
If the quarks were massless, $M_\pi$, $M_K$ and $M_\eta$ would vanish.
The masses of the Goldstone bosons$\,$\footnote{Sometimes, the name 
``Goldstone boson'' is 
reserved for the case of an exact symmetry, replacing it by the term  
``Pseudo-Goldstone-boson'' if the symmetry is an 
approximate one.} are due to the quark mass term in the
Hamiltonian of QCD. For all other multiplets, the main 
contribution to the mass is given by
the eigenvalue of $H_0$ --  the quark mass term $H_1$ only generates a small
perturbation that is responsible for the splitting of the levels, the state
with the largest strange quark component winding up at the top. For the
pseudoscalars, however, the main term is absent. First order perturbation
theory shows that the square of the pion mass is given by
$M_{\pi^+}^2=(m_u+m_d)B+\ldots\,$, where $B$ is the matrix element of 
$\ubar u$ in the unperturbed state. For the kaon, the leading term involves
the mass of the strange quark, $M_{K^+}^2=(m_u+m_s)B+\ldots\,$, 
$M_{K^0}^2=(m_d+m_s)B+\ldots\,$ The square of the kaon mass is about 13 times 
larger than the square of the 
pion mass because $m_s$ is about 13 times larger than $m_u+m_d$. Eightfold way
symmetry is perfectly consistent with
the fact that $m_u$, $m_d$, $m_s$ and hence $M_\pi$, $M_K$, $M_\eta$
are very different from one another. In this context, the
symmetry only implies that the matrix elements of the operators $\ubar u$,
$\dbar d$, $\sbar s$ in the various states of the pseudoscalar octet are
approximately determined by one and the same constant $B$. 

\section{Effective action}
I now turn to the Green functions of the various
operators built with the quark fields: vector or
axial currents,
as well as scalar or pseudoscalar densities. These are conveniently collected
in the effective action of the theory, which represents the response of the
system to the perturbation generated by corresponding external fields.
I denote the
external field associated with the vector current by $v_\mu (x)$
and extend the Lagrangian by the term $\bar{q}\gamma^\mu v_\mu (x) q$.
The field $v_\mu (x)$ is a matrix in flavour space, but is colour
neutral and also commutes with the Dirac matrices. Similarly, the axial
currents are generated by a term of the form $\bar{q} \gamma^\mu \gamma_5 a_\mu
(x)q$. To include the scalar and pseudoscalar densities, it suffices to
consider a space-dependent, complex quark mass matrix $m=m(x)$.
For reasons which will become clear shortly, it is convenient to
introduce a further external field $\theta (x)$, coupled to
the winding number density $\omega$ defined in eq.~(\ref{eq:omega}).
In the presence of these external fields, the QCD Lagrangian takes the
form
\bea
\label{wi21}
{\cal L}_{\QCD}\al=\al{\cal L}_{\ind G} +  \bar{q} i\gamma^\mu (\partial_\mu
-iG_\mu )q  + \bar{q}\gamma^\mu
(v_\mu +a_\mu \gamma_5)q -\bar{q}_{\mbox{\scriptsize R}}m\,
q_{\mbox{\scriptsize L}} -\bar{q}_{\mbox{\scriptsize L}}m^\dagger
q_{\mbox{\scriptsize R}}\no
{\cal L}_{\ind G} \al=
\al - \frac{1}{2g^2} \mbox{tr}_cG_{\mu\nu} G^{\mu\nu} - \theta\,
\omega \fs
\eea
The effective action of QCD is the logarithm of the corresponding
vacuum-to-vacuum transition amplitude,
\bea
\label{wi20}
\exp \,i\,S_{\eff}\{v,\,a,\,m,\,\theta  \} = \langle 0 \,\mbox{out}\mid 0\,
\mbox{in}\rangle\!\rule[-0.3em]{0em}{0em}_{\;v,\,a,\,m,\,\theta}
\eea
and contains the various external fields as arguments.
By construction, the quantities $v_\mu (x),\,a_\mu (x)$ and $m(x)$ are
$\mbox{N}_f\times \mbox{N}_f$ matrices acting in flavour space. While
$v_\mu (x)$ and $a_\mu (x)$ are hermitean, the field
$m(x)$ contains both a hermitean part, generating the scalar quark
densities, and an antihermitean part, giving rise to the pseudoscalar
operators.

The expansion of the effective action in powers of the external
fields $v_\mu,\,a_\mu,\,m$ and $\theta$
generates the Green functions of the
massless theory. The quark condensate, for instance, is
given by the term linear in $m(x)$, all other sources being switched
off,
\bea
\label{wi23}
S_{\eff} \al=\al
-\int\!\!d x\,\lvac\bar{q}_{\indR}m
q_{\indL}
+\bar{q}_{\indL}m^\dagger q_{\indR}
\rvac+\ldots
\eea
The various two-point functions are contained in the terms involving two
external fields. In particular, the correlation function of the axial
current is given by the term
\bea
\label{wi24}
S_{\eff} =\mbox{$\frac{1}{2}$}\,i\int\!\! d x d y\,
a_\mu^a(x)a_\nu^b(y)\lvac T A^\mu_a(x) A^\nu_b(y)\rvac
+\ldots \eea
where the field $a_\mu^b(x)$ represents the matrix $a_\mu (x)$
in the Gell-Mann basis,  $a_\mu (x)=\mbox{$\frac{1}{2}$}\lambda_b\,
a_\mu^b(x)$.

The same effective action also contains the Green functions of real QCD.
To extract these, one considers the infinitesimal
neighbourhood of the physical
quark mass matrix $m_0$ rather than the vicinity of the point $m = 0\,$: Set
$m(x) = m_0
+ \tilde{m}(x)$ and treat
$\tilde{m}(x)$ as an external field. The expansion of the effective action
in powers of $v_\mu,\,a_\mu,\,\tilde{m}$ and $\theta$ yields the Green
functions of the vector, axial, scalar and pseudoscalar currents and of the
operator $G_{\mu\nu}\tilde{G}^{\mu\nu}$ for the case of physical interest,
where the quark masses are different from zero.

In path integral representation, the effective action of QCD is given
by
\bea
\label{eq:path integral}
\exp \,i\,S_{\eff}\{v,\,a,\,m,\,\theta  \}
={\cal N}\!\!\int\!
[dG]\,\exp \left(\mbox{$ i\!\int \!d x\,{\cal L}_{\ind G}$}\right)
\;\mbox{det}\,D
\co\eea
where $D$ is the Dirac operator
\be
\label{wi25}
D \;=i\gamma^\mu \{\partial_\mu -i(G_\mu +v_\mu + a_\mu \gamma_5)\}
- m\mbox{$\frac{1}{2}$}(1-\gamma_5)-
m^\dagger \mbox{$\frac{1}{2}$}(1+\gamma_5) \co
\ee
and ${\cal N}^{-1}$ is the path integral for $v_\mu=a_\mu=m=\theta=0$.
In the present context, where the electroweak interactions are switched off,
there is a sharp distinction between
the colour field $G_\mu$ and the flavour fields $v_\mu,\,a_\mu$:
While the former is a dynamical variable, which mediates the strong
interactions and is to be integrated over in the path integral, 
the latter are classical auxiliary 
fields. 

\section{Chiral symmetry in terms of Green functions: Ward identities}
The construction of the effective theory relies on the symmetry properties
of the Green functions, more specifically, on the Ward identities obeyed by
these. The Ward identities can be expressed as a remarkably simple property of
the effective action: Disregarding the anomalies, $S_{\eff}$ is invariant under
local chiral rotations. Since this property plays a central role in the
following, I briefly show how it can be established. 

Formally, the QCD Lagrangian is invariant under local
U(3)$_{\indR}\times$U(3)$_{\indL}$ rotations of the quark fields,
\bea q_{\indR}(x)'=V_{\indR}(x)q_{\indR}(x)\co\hspace{2em} 
q_{\indL}(x)'=V_{\indL}(x)q_{\indL}(x)\co \hspace{2em}V_{\indR},
V_{\indL}\in \mbox{U(3)}\co \eea
provided the external fields are transformed accordingly:
\bea \label{eq:trafo}
r_\mu(x)'\al=\al V_{\indR}(x)r_\mu(x)V_{\indR}^\dagger-i\partial_\mu 
V_{\indR}V_{\indR}^\dagger\co\no
l_\mu(x)'\al=\al V_{\indL}(x)l_\mu(x)V_{\indL}^\dagger-i\partial_\mu 
V_{\indL}V_{\indL}^\dagger\co\\
m(x)'\al=\al V_{\indR}(x)m(x)V_{\indL}^\dagger\co\nonumber\eea
with $r_\mu=v_\mu+a_\mu$, $l_\mu=v_\mu-a_\mu$.
As is well-known, however, only the modulus of the determinant of the
Dirac operator is invariant under this operation -- the phase picks up a
change. The determinant is unique up to a local polynomial formed
with the gluon and external fields. The polynomial may be chosen such that
the determinant is invariant under the subgroup generated by the
vector charges, as well as under gauge transformations of the gluon field.
Under the infinitesimal transformation
\bea V_{\indR}={\bf 1}+i\alpha+i\beta+\ldots\co
\hspace{2em}V_{\indL}={\bf 1}+i\alpha-i\beta+\ldots\co\nonumber\eea
the change in the phase of the determinant then takes the form
\bea\label{eq:det} \delta \ln \det D=-2i\int\!\!d x\,\langle\beta(x)
\rangle\,
\omega(x)-i\int\!\!d x\,\langle\beta(x)\,\Omega(x)\rangle\co\eea
where $\langle X\rangle$ denotes
 the trace of the $3\times 3$ matrix $X$.
The gluon field only enters the first term, through the winding number density
$\omega$ -- this term gives rise to the U(1)-anomaly in the conservation law
(\ref{eq:omega}) for the singlet axial current. The second term only
contains the external vector and axial fields,
\bea\label{eq:Omega} \Omega=\frac{N_c}{4\pi^2}\,\epsilon^{\mu\nu\rho\sigma}
\partial_\mu v_\nu\partial_\rho v_\sigma+\ldots\eea
The particular contribution indicated is the one that describes the
the anomalies in the Ward identities 
for the correlation function
$\lvac T A^\lambda V^\mu V^\nu\rvac$, which play a central role in the decay
$\pi^0\rightarrow\gamma\gamma$. The full
expression for $\Omega$ also contains terms that are quadratic in $a_\mu$, 
as well as contributions
involving three or four vector or axial fields, which account for 
the anomalies in the Ward identities obeyed by the 4- and 5-point functions. 
The explicit expression for these terms is not relevant in our context, but 
it is essential that they are independent of the gluon field: This property
implies that
the second term in eq.~(\ref{eq:det}) can be pulled out of the path integral.
The first one can be absorbed with a change in the vacuum angle -- this is the
reason for introducing a term proportional to $\omega$ in the definition of 
the effective action. The net result is that the change 
generated by an infinitesimal chiral rotation of the external fields,
\bea
\label{wi35}
\delta v_\mu \al=\al \partial_\mu \alpha + i[\alpha,v_\mu ] +i[\beta,a_\mu ]
\hspace{1em},\hspace{1em}
\delta a_\mu = \partial_\mu \beta + i[\alpha,a_\mu ] +i[\beta,v_\mu ] \no
\delta m\al=\al i(\alpha +\beta)m-im(\alpha -\beta)
\hspace{1em},\hspace{1em}
\delta \theta = -2\langle\beta\rangle\co
\eea
can be given explicitly: The effective action picks up the change
\be
\label{eq:anomalies}
\delta S_{\eff}\{v,\,a,\,m,\,\theta  \} = -\int\!\! d x\,
\langle\beta(x)\Omega(x)\rangle\fs
\ee
The relation collects all of the Ward identities obeyed by the Green
functions formed with the operators $\qbar\gamma_\mu \lambda q$,
$\qbar\gamma_\mu \gamma_5\lambda q$, $\qbar\lambda q$,
$\qbar \,i \gamma_5 \lambda q$ and $\omega$. 
It states that the effective action is gauge invariant under
local chiral rotations, except for the anomalies, which are
of purely geometric nature: The right-hand side of eq.~(\ref{eq:anomalies})
is independent of the coupling constant and of the quark masses --
it only involves the number $N_c$ of colours.

The relation $\delta \theta=-2\langle \beta\rangle$ specifies the
transformation law of the vacuum angle only for infinitesimal chiral
rotations. The one relevant for finite transformations is obtained by
integrating this relation. As the group U(1) is not simply connected,
the result, however, is unique only up to multiples of $2\pi$, so that only 
the transformation law for $e^{i\theta}$ is free of ambiguities:
\bea\label{eq:trafo theta} 
e^{i\theta'}=\det (V_{\indR}^\dagger V_{\indL})\, e^{i\theta}\fs\eea

\section{Basic low energy constants}
\label{basic}
Let us return to paradise and consider the quark condensate 
$\lvac \qbar_{\indR}\!\rule{0em}{0em}^\alpha  
q_{\indL}\!\rule{0em}{0em}^\beta\rvac$ where $\alpha,\beta=1,2,3$
indicates the quark flavour. Under left-handed chiral rotations, the operator
$\qbar_{\indR}\!\rule{0em}{0em}^\alpha  
q_{\indL}\!\rule{0em}{0em}^\beta$ transforms according to the
representation $\underline{3}$.
If the ground state were invariant under these, the condensate would 
therefore vanish. In this sense, the matrix element 
$\lvac \qbar_{\indR}\!\rule{0em}{0em}^\alpha  
q_{\indL}\!\rule{0em}{0em}^\beta\rvac$ represents a quantitative
measure for the strength of the spontaneous symmetry breakdown and is referred
to as an order parameter. The vacuum expectation value of any scalar
operator that is not invariant under chiral transformations may serve as an
order parameter. The quark condensate is the most
important one, because it represents the one of lowest dimension.
As there is no reason for the condensate to vanish, one generally assumes
that it is different from zero. Lattice calculations provide some evidence for
this to be the case, but it is notoriously difficult to explore the 
properties of the theory for small quark masses, not to speak of those of the
massless theory.

Since the ground state is invariant 
under SU(3)$_{\indV}$, the condensate involves a single constant,
\bea\label{eq:condensate} \lvac \qbar_{\indR}\!\rule{0em}{0em}^\alpha  
q_{\indL}\!\rule{0em}{0em}^\beta\rvac=-\mbox{$\frac{1}{2}$}
\delta^{\alpha\beta}\,C\fs\eea 
Moreover, invariance of the ground state under space reflections implies that
$\lvac \qbar\, i \gamma_5  q\rvac$ vanishes, so that C is real.

A nonzero condensate immediately implies that the spectrum contains massless
particles. To see this, consider the correlation function
of the axial and pseudoscalar current octets (to distinguish the
octet components from the singlets, I label these with the indices
$i,k=1,\,\ldots\,,\,8$, while $a,b=0,\,\ldots\,,\,8$)
\bea A^\mu_i=\qbar \gamma_\mu\gamma_5\mbox{$\frac{1}{2}$}
\lambda_i q\co\hspace{2em}P_i=\qbar \,i \gamma_5 \mbox{$\frac{1}{2}$}
\lambda_i q\co\nonumber\eea
which obeys the Ward identity
\bea\label{eq:Ward identity} \partial_\mu \lvac T A^\mu_i(x) P_k(0)\rvac=
-\mbox{$\frac{1}{4}$} i\,  \delta(x)
\lvac \qbar \{\lambda_i,\lambda_k\} q\rvac\fs\eea
The relation may be derived in a formal manner, by evaluating the derivative
of the time-ordered product and using the equation of motion for the quark
fields, but this is not without caveats -- in general, operations of this sort
are afflicted by ambiguities. The result may however be established in full
rigour from the effective action. It suffices to consider the terms
\bea \al\al S_{\eff}=-\int\!\!d x\,\lvac\qbar_{\indR}m q_{\indL}+
\qbar_{\indL} m^\dagger q_{\indL}\rvac
\no\al\al -i\!\int\!\!d xd y \lvac T \{\qbar \gamma_\mu\gamma_5 
a^\mu q\}_x\, \{\qbar_{\indR}m q_{\indL}
+\qbar_{\indL} m^\dagger q_{\indR}\}_y\rvac +\ldots
\nonumber\eea
and to apply the infinitesimal chiral rotation (\ref{eq:trafo}) to the
external fields $a_\mu(x)$, $m(x)$. The relation (\ref{eq:anomalies}) requires
the changes proportional to $\{\beta, m\}$, $\{\beta, m^\dagger\}$ in the first
line to cancel those proportional to $\partial_\mu \beta(x) m(y)$,  
$\partial_\mu \beta(x) m^\dagger(y)$ in the second line. This
condition indeed leads to eq.~(\ref{eq:Ward identity}).

The above Ward identity can be solved explicitly. Lorentz
invariance implies that the Fourier transform is of the form
\bea \int\!\! dx \,e^{i p\cdot x}\, \lvac T A^\mu_i(x) P_k(0)\rvac = 
p^\mu\Pi_{ik}^{\mbox{\tiny AP}}(p^2)\fs\nonumber\eea
With the explicit representation
(\ref{eq:condensate}) for the condensate, the identity
(\ref{eq:Ward identity}) thus becomes
\bea - p^2 \Pi_{ik}^{\mbox{\tiny AP}}(p^2)= \delta_{ik}C\fs\eea
Hence the function $\Pi_{ik}^{\mbox{\tiny AP}}(p^2)$ contains a pole at 
$p^2=0$ -- the spectrum must contain massless particles. The result also shows
that the correlation function under consideration is fully determined by
the condensate:
\bea \int\!\! dx \,e^{i p\cdot x}\,\lvac T A^\mu_i(x) P_k(0)\rvac
 =\delta_{ik}\frac{p^\mu C}{-p^2-i\epsilon}\fs\eea
The pole requires the existence of 8 massless one-particle states 
$|\pi^i\rangle$, with nonzero matrix elements 
\bea \lvac A^\mu_i|\pi^k\rangle=i\, \delta^k_i\,p^\mu\,F\co\hspace{2em}
\lvac P_i|\pi^k\rangle= \delta^i_k\,G\co\hspace{2em} F\,G=C\fs\eea
In fact, only these intermediate states contribute and it is easy to see why 
that is so. The matrix element $\langle n| P_k\rvac$ vanishes unless the
angular momentum of the state $\langle n|$ vanishes. For such states, however,
Lorentz invariance implies that the matrix element $\lvac A^\mu_i |n\rangle$
is proportional to the momentum $p^\mu_n$ of the state. At the same time,
current conservation requires the matrix element to be transverse to
$p^\mu_n$. The two conditions can only be met if $p_n^2=0$, that is, if the
vector $|n\rangle$ describes a massless particle of spin zero -- a Goldstone
boson.  

The calculation demonstrates the validity of the Goldstone theorem in the
context of QCD: The massless theory can have a nonzero quark 
condensate only if (a) the spectrum contains Goldstone bosons and (b)
the corresponding one particle matrix elements of the axial and pseudoscalar
currents are different from zero. The value of the 
pion matrix element of the axial current, the pion decay constant
$F_\pi=92.4\,\mbox{MeV}$, is known experimentally, from 
the decay $\pi\rightarrow\mu\nu$. Since the Goldstone bosons of QCD
are the pseudoscalar mesons, this constant
must approach a nonzero limit $F$ when the quark masses are sent to zero. 

The Gell-Mann-Oakes-Renner
relation$\,$\cite{Gell-Mann Oakes Renner} is an immediate consequence: 
For nonzero quark masses, the divergence
of the axial current is related to the pseudoscalar density by
$\partial_\mu (\ubar \gamma^\mu\gamma_5 d)=(m_u+m_d)\,\ubar\,i \gamma_5 d$.
The vacuum-to-pion matrix element of this equality shows that the
physical one particle matrix elements obey the exact
relation $M_\pi^2 F_\pi=(m_u+m_d) G_\pi$. The coefficient of the leading term
in the expansion of the pion mass in powers of the quark masses,
$M_\pi^2=(m_u+m_d) B+\ldots\,$, is therefore determined by the condensate
and by the pion decay constant: $B=G/F=C/F^2$. 

The above relations involve two independent low energy constants, $F$ and $B$.
In particular, the value of the quark condensate in the massless
theory may be expressed in terms of these: $\lvac \ubar u\rvac=-F^2 B$.
In fact, the same two constants fully determine the leading low energy
singularities in all of the Green functions. In the case of the
correlation function $\lvac T A^\mu_i(x) A^\nu_k(0)\rvac$, for instance,
this can be seen as follows.  
In the massless theory, this function obeys the Ward identity
$\partial_\mu\lvac T A^\mu_i(x) A^\nu_k(0)\rvac=0$, so that the Fourier
transform is transverse to the momentum,
\bea\label{eq:PiAA} i\!\int \!\!dx\,e^{i p\cdot x}\,\lvac T A^\mu_i(x) 
A^\nu_k(0)\rvac=
(p^\mu p^\nu-g^{\mu\nu} p^2)\,\delta_{ik}\Pi^{\mbox{\tiny AA}}(p^2)\fs\eea
The one-particle intermediate states generate a pole at $p^2=0$, 
whose residue is also determined by the constant $F$:
\bea\label{eq:pole term} 
\Pi^{\mbox{\tiny AA}}(p^2)=\frac{F^2}{-p^2-i\epsilon}+\ldots\eea

\section{Low energy expansion of the effective action}

The result obtained in the preceding section for the quark condensate
and for the correlation functions of the operators $A^\mu_i$, $P_i$ 
amounts to the following explicit expression for the relevant terms in 
the effective action:
\bea\al\al\label{eq:expansion of Seff} 
S_{\eff}=\int\!\!  dx\,\mbox{$\frac{1}{2}$}F^2 B\, \langle
m(x)+m^\dagger(x)\rangle\no\al\al - 
\int\!\!dx dy\,\mbox{$\frac{1}{2}$}i F^2 B \,\sum_{i=1}^8 
\partial^\mu a_\mu^i(x)\, \Delta_0(x-y)\, 
\langle \lambda_i(m(y)-m^\dagger(y)\rangle \\\al\al-
\int\!\!dx dy\,\mbox{$\frac{1}{4}$} F^2  \,\sum_{i=1}^8a_{\mu\nu}^i(x)
\Delta_0(x-y)  
a^{\mu\nu\,i}(x)+\ldots\nonumber
\eea
The function $\Delta_0(x)$ stands for the massless propagator,
\bea \Delta_0(x)=\frac{1}{(2\pi)^{4}}\!\int \!\!dp\,\frac{e^{-i p\cdot
    x}}{-p^2-i\epsilon}=\frac{i}{4\pi^2(-x^2-i\epsilon)}\co\nonumber\eea 
and $a_{\mu\nu}^i$ is the field strength, 
$a_{\mu\nu}^i=\partial_\mu a_\nu^i-\partial_\nu a_\mu^i$.

The expression shows that the effective action of massles QCD is an inherently
nonlocal object -- the exchange of Goldstone bosons 
implies that the correlation functions only drop off with a power of the
distance. This is in marked contrast to the effective action of Heisenberg 
and Euler, who considered the
correlation functions of the current for electrons exposed to an external
electromagnetic field.$\,$\cite{Heisenberg Euler} 
The corresponding effective action is given by the sum 
of all one-loop
graphs containing an arbitrary number of external field vertices,
\bea\exp\, i\, S_{\eff}^e\{A\}=\det 
\left\{i\gamma^\mu(\partial_\mu - i\, e A_\mu)-m_e\right\}
\fs\eea 
In this case, the correlation functions drop off exponentially with the 
distance. If the external field is weak $(|F_{\mu\nu}|\ll m_e^2,\,
F_{\mu\nu}\equiv\partial_\mu
A_\nu-\partial_\nu A_\mu$) and
varies only slowly, so that the typical
wavelengths are small compared to the Compton wavelength of the electron
$(|\partial_\lambda F_{\mu\nu}|\ll m_e | F_{\mu\nu}|)$, the effective action
may be expanded in powers of derivatives:
\bea  \al\al S_{\eff}^e\{A\}= \int\!\! dx\, {\cal L}_{\eff}\co\no
\al\al {\cal L}_{\eff}=-\frac{e^2\Pi^{\hspace{0.05em}\mbox{\tiny 
j\hspace{0.05em}j}}(0)}{4} 
F^{\mu \nu }\,F_{\mu \nu } -
\frac{e^2}{240\pi^2m_e^2}\;\partial^\lambda F^{\mu
\nu }\,\partial_\lambda F_{\mu \nu } \\\al\al-
 \frac{e^2}{2240 \pi^2m_e^4}\;
\raisebox{0.2em}{\fbox{\rule[0.15em]{0em}{0em}\,}}\,F^{\mu \nu}
\raisebox{0.2em}{\fbox{\rule[0.15em]{0em}{0em}\,}}\,F_{\mu \nu}
+\frac{e^4}{1440\pi^2m_e^4}\{(F^{\mu \nu }\,F_{\mu \nu
})^2 + 7(F^{\mu \nu }\,\tilde{F}_{\mu \nu })^2\}
+\ldots\nonumber \eea
The contributions quadratic in the field strength are described by the 
vacuum polarization $\Pi^{\mbox{\tiny 
j\hspace{0.1em}j}}(p^2)$ -- the analogue of the 
quantity $\Pi^{\mbox{\tiny AA}}(p^2)$ introduced in eq.~(\ref{eq:PiAA}), 
the axial current being replaced by the
electromagnetic one. The various terms arise from the Taylor series expansion
of this function in powers of
$p^2/m_e^2$. The first term in the Heisenberg-Euler Lagrangian 
is proportional to the Lagrangian of the free
electromagnetic field and merely renormalizes the photon wave function,
$A_\mu^{\ren} = \{1 + \frac{1}{2}e^2\Pi^{\mbox{\tiny 
j\hspace{0.1em}j}} (0)\}A_\mu $. The remainder
amounts to a modification of the Maxwell Lagrangian and deforms the
electromagnetic field generated by a given charge distribution. For
slowly varying fields, the effect is dominated by the contribution of order
$(\partial_\lambda F_{\mu \nu })^2$, referred to as the Uehling term. In
particular, this
term generates a small contribution to the Lamb shift (spacing
between the $S$-- and $P$--wave
bound states of the hydrogen atom with principal quantum number $n=2$).

The origin of the qualitative difference between the two effective actions is
evident: The spectrum of the states that can be created by
an external electromagnetic field has an energy gap, $\Delta E=2 m_e$, while
the spectrum of massless QCD does not. External fields can generate Goldstone
bosons, even if their wavelength is large. In reality, QCD also has
an energy gap: $\Delta E=M_\pi$. Accordingly, for external fields that vary
only slowly on the scale set by the Compton wavelength of the pion, the 
effective action 
also admits a derivative expansion that consists of a string of local
terms. The range of validity of such a representation, however, is
very limited, because $m_u,\,m_d$ and hence the pion mass are small. 
If the quark masses are set equal to zero, the straightforward 
derivative expansion becomes entirely meaningless.

In contrast to the case of the correlation function $\lvac T A^\mu_i
P_k\rvac$, which exclusively receives a contribution from the exchange of 
single Goldstone bosons, the quantity 
$\lvac T A^\mu_i A^\nu_k\rvac$ also picks up contributions
from spin 1 intermediate states with $3,5,\ldots$ pseudoscalar mesons, which
the representation for the effective action in eq.~(\ref{eq:expansion of
  Seff})  does not account for.  
In the function  $\Pi^{\mbox{\tiny AA}}(p^2)$, these 
generate a branch cut starting at $p^2=0$. At low energies, phase space 
strongly suppresses the discontinuity across the cut, but in the
vicinity of the resonance $a_1(1260)$, there is a pronounced peak.
The pole term dominates in the sense that the remainder approaches a finite 
limit $H$ when $p\rightarrow 0$. We may view the pole
as the leading term in the expansion of $\Pi^{\mbox{\tiny AA}}(p^2)$ in powers
of the momentum:
\bea\label{eq:two terms} 
\Pi^{\mbox{\tiny AA}}(p^2)=\frac{F^2}{-p^2-i\epsilon}+H+O(p^2)\fs\eea
The constant $H$ gives rise to an additional contribution in the expression
for the effective action: a local term proportional to $\int\! dx\,H\sum_i
(a_{\mu\nu}^i)^2$, which resembles the leading term in the Heisenberg-Euler
action. The systematic analysis of the effective theory, which will be
sketched below, orders the contributions according to powers of the momentum
and automatically accounts for the above term at first nonleading order
of this ``low energy expansion''. The imaginary
part generated by intermediate states with 3 Goldstone bosons 
manifests itself at next-to-next-to leading order of that 
expansion. The resulting representation for the
effective action in effect also amounts to 
a derivative expansion. It is of a nonlocal type, because the
expansion contains inverse powers of the momentum.

The pole term provides an adequate approximation of the function 
$\Pi^{\mbox{\tiny AA}}(p^2)$ only for
momenta that are small compared to the mass of the $a_1(1260)$. In the case of
the correlation function of
the vector current, where the $\rho(770)$ generates a peak in the imaginary
part, the domain where the first one or two terms in the expansion in powers 
of the momentum provide a decent approximation is even smaller. Generally
speaking, all of the momenta must be small compared to the intrinsic scale of 
QCD. The quantitative form of this condition depends on the 
channel under study. The internal consistency of the effective theory
also leads to a constraint on the magnitude of the momenta:
As will be discussed below, the loop graphs generate contributions at 
next-to-leading order. These must be small
compared to the leading terms, a requirement that only holds
if the momenta obey the condition $p\ll 4 \pi F_\pi/\sqrt{N_f}\simeq
700\,\mbox{MeV}$. 
 
\section{Effective Lagrangian}

The expression (\ref{eq:expansion of Seff}) is reminiscent of the effective 
action of a free field theory
that involves massless scalar fields. Indeed, we can introduce eight
pseudoscalar fields $\pi^1(x),\ldots\,,\,\pi^8(x)$ and consider the Lagrangian
\bea\label{eq:Leff2} {\cal L}_{\eff}\al=\al\mbox{$\frac{1}{4}$}\langle  
d_\mu \pi d^\mu\pi\rangle+\mbox{$\frac{1}{2}$}F^2 B\, \langle
m+m^\dagger\rangle -\mbox{$\frac{1}{2}$}i FB\langle \pi\, m-\pi\, 
m^\dagger\rangle\co\no
\pi\al=\al\sum_{i=1}^8\pi^i\lambda_i\co\hspace{2em} d_\mu \pi=\sum_{i=1}^8
(\partial_\mu\pi^i-F a_\mu^i)\lambda_i\co\eea
which is quadratic in these fields, so that the effective action 
coincides with the classical action. The corresponding equation of motion
reads:
\bea \wave\pi^i(x)=F\partial^\mu a_\mu^i(x) -iFB\langle\lambda_i \{
m(x)-m^\dagger(x)\}\rangle\fs\eea
It is straightforward to work out the classical action of this model
and to check that the result of this calculation within classical field theory
does reproduce the terms in eq.~(\ref{eq:expansion of Seff}). The same
calculation also yields the pole term in the correlation 
function $\lvac TP_i(x) P_k(0)\rvac$, which is not included there. 

The part of the effective action of massless QCD that we have picked out may
thus equally well be described 
in terms of an entirely different field theory, which instead of quarks and
gluons contains a set of pseudoscalar fields as dynamical variables. 
In contrast to the effective action, the relevant Lagrangian does represent a
local expression. It is convenient to count the external field 
$a^i_\mu(x)$ as a quantity of the same order as the derivative, 
$O(a)=O(\partial)=O(p)$ and to book the field $m(x)$ as a term of $O(p^2)$.
The above expression for the Lagrangian then represents a term of $O(p^2)$.

It is
clear, however, that this framework is incomplete: (a) it does not cover all
of the Green functions of QCD and (b) the corresponding representation for the
correlation function $\lvac T A^\mu_i(x) A^\nu_k(0)\rvac$ only accounts for
the contribution that dominates at low energies. As discussed below, 
both of these limitations can
be removed, at least in principle: If the effective Lagrangian is chosen
properly, the resulting effective action coincides with the one of QCD,
to any desired order in the low energy expansion. In this sense, there
exists an alternative, exact representation of QCD: The path
integral in eq.~(\ref{eq:path integral}) can be replaced by a path integral
over the effective fields $\pi^1(x),\ldots\,,\,\pi^8(x)$,
\bea \label{eq:effective path integral} 
\exp \,i\,S_{\eff}\{v,a,m,\theta\}={\cal N}_{\eff}\!\int\!
[d\pi]\,\exp \left(\mbox{$ i\!\int\! dx\,$}{\cal L}_{\eff}\right) \co\eea
where ${\cal N}_{\eff}^{-1}$ is the same path integral evaluated at 
$v_\mu=a_\mu=m=\theta=0$.
The full effective Lagrangian is a local expression of the form
\bea {\cal L}_{\eff}={\cal L}_{\eff}(\pi,v,a,m,\theta;\partial\pi,\partial
v,\partial a,\partial m,\partial \theta;\partial^2\pi,\ldots)\fs\nonumber
\eea
It can be ordered by counting the number of fields and derivatives with
\bea \{\pi,\theta\}=O(1)\co\hspace{2em}
\{\partial,v,a\}=O(p)\co\hspace{2em}m=O(p^2)\fs\eea
Lorentz invariance permits only even
orders. The expansion starts at $O(p^2)$:
\bea {\cal L}_{\eff}={\cal L}^{(2)}+{\cal L}^{(4)}+
{\cal L}^{(6)}+\ldots\eea

For a detailed proof of this claim, I refer to the 
literature.$\,$\cite{foundations} In the following, I first briefly
describe the main properties of the effective Lagrangian and of the  
path integral (\ref{eq:effective path integral}) and then make a few comments
about the proof of the statement that the effective theory 
reproduces the Green functions of QCD, order by order in the low energy 
expansion.

\section{Effective field theory}
The leading term in the derivative expansion of ${\cal L}_{\eff}$
is the Lagrangian of the nonlinear $\sigma$-model:
\bea {\cal L}^{(2)}=
\mbox{$\frac{1}{4}$}\,F^2\langle \nabla_\mu U \nabla^\mu U^\dagger\rangle
+\mbox{$\frac{1}{2}$}\,F^2B\,\langle m\, U^\dagger+Um^\dagger\rangle+
\mbox{$\frac{1}{12}$} H_0 D_\mu\theta D^\mu\theta\fs\nonumber\eea
The effective field is described in terms of a unitary $3\times 3$ matrix,
$U U^\dagger ={\bf 1}$. The covariant derivatives stand for
\bea \nabla_\mu U\al=\al\partial_\mu U-i(v_\mu+a_\mu)
U +i\, U(v_\mu-a_\mu)\,,\hspace{1em}
D_\mu\theta =\partial_\mu\theta+2\langle
a_\mu\rangle \,.\nonumber\eea

With the parametrization
$U(x)=\exp\,i\,\pi(x)/F$, one readily
checks that ${\cal L}^{(2)}$ indeed contains the effective
Lagrangian given in eq.~(\ref{eq:Leff2}): That part accounts for the 
first three terms in the expansion 
in powers of the field $\pi(x)$, for $v_\mu=\langle a_\mu\rangle=\theta=0$.
The expansion does not stop there, however. The 
nonlinear $\sigma$-model Lagrangian also contains vertices describing
interactions among the Goldstone bosons, and contributions
that involve the external fields $v_\mu,\langle a_\mu\rangle$ and $\theta$: 
This Lagrangian also accounts for the leading terms in the low energy expansion
of the scattering amplitudes and of the 
correlation functions of the operators $V^\mu_a$, $A^\mu_0$, $\omega$. 
In fact, all of the predictions obtained in the sixties,
on the basis of current algebra and PCAC, can be worked out in a comparatively
very simple manner with this Lagrangian. Apart from the term $H_0
D_\mu\theta D^\mu\theta$, which exclusively generates a contact 
contribution in the two-point functions of the operators $A^\mu_0$ and
$\omega$, the Lagrangian ${\cal L}^{(2)}$ only involves the two basic low
energy constants $F$ and $B$ introduced in section \ref{basic}. In the large
$N_c$ limit, $F$ is of order $\sqrt{N_c}$, while B and $H_0$ are of order
1.

The crucial property that distinguishes the nonlinear $\sigma$-model 
from all other field theory models with eight
scalar or pseudoscalar fields as dynamical variables is that it is manifestly
invariant 
under the group  U(3)$_{\indR}\times$U(3)$_{\indL}$ of local chiral rotations: 
The transformation (\ref{eq:trafo}), (\ref{eq:trafo theta}) of the external
fields leaves the Lagrangian ${\cal L}^{(2)}$ invariant, provided the meson
field $U(x)$ is transformed with
\bea\label{eq:trafo U} U(x)'=V_{\indR}(x)U(x) V_{\indL}^\dagger(x)\fs\eea
Note that this transformation law does not leave the determinant of $U(x)$
invariant. Indeed, $U(x)$ is an element of SU(3) only for $\theta=0$.
The $\theta$-term in the Lagrangian of QCD -- 
which is needed to analyze the consequences of chiral symmetry for the full 
group U(3)$_{\indR}\times$U(3)$_{\indL}$ of chiral rotations --
modifies the condition $\det U=1$ that pertains to the standard nonlinear
$\sigma$-model: The condition is replaced by the constraint
\bea \det U=e^{-i\theta}\co\eea
which, in view of eq.~(\ref{eq:trafo theta}), is consistent with the 
transformation law (\ref{eq:trafo U}). 

I add a remark of technical nature. In the presence of the singlet external
fields $\theta$ and $\langle a_\mu\rangle$, the covariant derivative 
$\nabla_\mu U$ is not convenient to work with, because the trace 
$\langle U^\dagger\nabla_\mu U\rangle$ does not vanish. In the following,
I instead use  $D_\mu U=\nabla_\mu
U+\frac{1}{3} i D_\mu\theta\, U$,  which does obey
$\langle U^\dagger D_\mu U\rangle=0$. In this notation, 
the leading term of the effective Lagrangian takes the form
\bea {\cal L}^{(2)}\al=\al
\mbox{$\frac{1}{4}$}\,F^2\langle D_\mu U D^\mu U^\dagger\rangle
+\mbox{$\frac{1}{2}$}\,F^2B\,\langle m\, U^\dagger+Um^\dagger\rangle+
\mbox{$\frac{1}{12}$}\tilde{H}_0 D_\mu\theta D^\mu\theta\co\no
D_\mu U\al=\al\partial_\mu U-i\,(v_\mu+a_\mu)
U +i\, U(v_\mu-a_\mu)+\mbox{$\frac{1}{3}$}\,i
\,D_\mu\theta\, U\co\eea
with $\tilde{H}_0= H_0+F^2$. Note that, in the large $N_c$ limit, 
$\tilde{H}_0$ is of $O(N_c)$ -- the leading term in the $1/N_c$ expansion of
this constant is given by $F^2$.

The invariance of ${\cal L}^{(2)}$ immediately implies
that the corresponding classical action is invariant under local chiral
transformations of the external fields. The classical action collects the set
of all tree graph contributions to the path integral, so that this part of the
effective action is invariant. The loop graphs
cannot simply be dropped -- otherwise, unitarity is violated -- but they
represent contributions of nonleading order:$\,$\cite{Weinberg 1979}
In dimensional regularization, those graphs of ${\cal L}^{(2)}$ that
contain $\ell$ meson loops represent contributions of order 
$p^{2\ell +2}$. Accordingly, at leading order of the low energy
expansion, only the tree graphs matter. The claim that the effective
action of QCD can be represented as a path integral over meson fields
thus implies that the leading contributions in the expansion of this effective
action are given by the classical action of the nonlinear $\sigma$-model and
are therefore invariant under chiral rotations. This conclusion is
in agreement with the fact that the anomalous terms occurring in the
Ward identities represent contributions of order $p^4$: As indicated by
the expression in eq.~(\ref{eq:Omega}), the quantity $\Omega$ represents a
term of that order. 

\section{Illustration: Topological susceptibility}

At low energies, 
the leading contributions to the Green functions of QCD are given by
the tree graphs of ${\cal L}^{(2)}$. I illustrate the content of this 
claim with
the topological susceptibility,$\,$\footnote{The sign convention adopted here
is such that the imaginary part of $\chi(q^2)$ is positive. In this
convention, $\chi(0)$ is negative: The mean square
winding number per unit volume of euclidean space is given by $-\chi(0)$.}  
\bea\label{eq:ts} \chi(q^2)=i\!\int \!\!dx e^{iq\cdot x} 
\lvac T \,\omega(x)\,\omega(0)\rvac\fs\eea 
Because the dependence of the susceptibility on the quark masses 
is very interesting, I do not set $m_u$ equal to $m_d$.
To evaluate the corresponding low energy representation at leading 
order, it suffices to calculate the extremum of the classical action
of ${\cal L}^{(2)}$ in the presence of the external field $\theta(x)$,
while $m(x)$ is identified with the 
physical quark mass matrix and all other external fields are switched
off. The correlation function of interest is the coefficient of the term
quadratic in $\theta(x)$. The classical equation of motion implies that
all components of $\pi(x)$ vanish at the extremum, except $\pi^3(x)$
and $\pi^8(x)$. The quadratic part of the Lagrangian yields the corresponding
masses. The states
$\pi^0$ and $\eta$ mix and the levels repel. The eigenvalues are
\bea \al\al M^2_{\pi^0}= (m_u+m_d)B-\Delta\co\hspace{2em}
M^2_{\eta}=\mbox{$\frac{2}{3}$}\,(\hat{m}+2 m_s) B+\Delta\co\no
\al\al\Delta= \frac{4\sin^2\!\epsilon }{3\cos2\, \epsilon}\,(m_s-\hat{m})B\co\,
\hspace{2em}\mbox{tg}\,2\,\epsilon=\frac{\sqrt{3}}{2}\,
\frac{m_d-m_u}{m_s-\hat{m}}\fs\nonumber\eea
The susceptibility is obtained by solving the classical
equation of motion to first order in $\theta$. The result 
reads:$\,$\cite{GL 1985}
\bea 
\al\al\chi(q^2)= \sum_{P=\pi^0,\eta}\frac{|\lvac\omega|P\rangle|^2}
{M_P^2-q^2}-\mbox{$\frac{1}{9}$}\,B F^2(m_u+m_d+m_s) +
\mbox{$\frac{1}{6}$}\, \tilde{H}_0\, q^2+O(p^4)\co\no
\al\al \lvac \omega
|\pi^0\rangle=\frac{1-\frac{4}{3}\sin^2\!\epsilon}{2\cos\epsilon}\,
(m_d-m_u) B F\co\no
\al\al\lvac \omega |\eta\rangle=\frac{2(1-4\sin^2\!\epsilon)\cos\epsilon}
{3\sqrt{3}\,\cos2\,\epsilon}\,(m_s-\hat{m}) B F\fs
\nonumber\eea

The quantity $\chi(0)$
represents the second derivative of the vacuum energy with respect to $\theta$
and must vanish if either $m_u$, $m_d$ or $m_s$ are sent to zero,
because the $\theta$-dependence then disappears.$\,$\cite{Crewther} 
Indeed, the pole contribution in the above representation cancels the momentum 
independent term if one of the quark masses is turned off. The result takes
the simple form:$\,$\cite{Di Vecchia Veneziano}
\bea \chi(0)=- B F^2 m_{red}+O(m^2)\co\eea
where $m_{red}$ stands for the reduced mass, 
\bea
\frac{1}{m_{red}}=\frac{1}{m_u}+\frac{1}{m_d}+\frac{1}{m_s}\fs\eea
The relation amounts to a low energy theorem for the mean square winding number
per unit volume: The expansion of this quantity in powers of the quark masses 
starts with $\langle \nu^2\rangle/V=B F^2 m_{red}+O(m^2)$. 

The first derivative
$\chi'(0)$ is of interest in connection with the spin content of the
proton.$\,$\cite{Shore Veneziano,Ioffe} The explicit expression 
reads: 
\bea\label{eq:chiprim} \chi'(0)=\mbox{$\frac{1}{2}$} F^2 m_{red}^2
\left\{\frac{1}{m_u^2}+\frac{1}{m_d^2}+
\frac{1}{m_s^2}\right\}+\mbox{$\frac{1}{6}$}\,H_0+O(m)\fs\eea
As mentioned earlier, the second term is suppressed as compared to the first:
$F^2=O(N_c)$, $H_0=O(1)$. In the following, I only consider the leading 
contribution, which is fully determined by the quark mass ratios $m_u:m_d:m_s$.

Numerically, inserting the phenomenological values of the mass
ratios,$\,$\cite{quark mass ratios} and using $F\simeq F_\pi=92.4
\,\mbox{MeV}$, the formula predicts
$\chi'(0)=2.2 \cdot 10^{-3}\,\mbox{GeV}^2$, in remarkable agreement with
the sum rule determination of Ioffe and collabo\,ra\-tors,$\,$\cite{Ioffe} 
who find
$\chi'(0)=(2.3\pm 0.6) \cdot 10^{-3}\,\mbox{GeV}^2$ and
$(2.0\pm 0.5) \cdot 10^{-3}\,\mbox{GeV}^2$, depending on the method used. 
Stated differently, the sum rule results confirm the validity of 
the Okubo-Iizuka-Zweig rule in this case: The contribution from the term
$H_0$ is numerically small. 

Note that the above algebraic result is
extremely sensitive to the pattern of chiral symmetry breaking:
In view of $m_u,m_d\ll m_s$, the formula roughly yields 
$\chi'(0)\simeq \frac{1}{2}F^2(m_u^2+m_d^2)/
(m_u+m_d)^2$, so that the result is nearly independent of $m_s$, 
but changes by a factor of 2 if the ratio $m_u/m_d$ is varied between 0 and 1.
The leading term in the quark mass expansion of $\chi'(0)$ is of a similar
structure as the one in the mass ratio
$(M_{K^0}^2-M_{K^+}^2)/M_\pi^2\simeq (m_d-m_u)/(m_u + m_d)$.
The connection with the proton spin content indicates that a similar
sensitivity to the ratio $m_u/m_d$ also occurs in the relevant nucleon
matrix elements.

In principle, one could also derive the above results by means of current 
algebra and PCAC, but the calculation would be considerably more tedious. 
Beyond leading order, it is practically impossible to study such quantities 
without making use of the effective Lagrangian method.

I add a comment$\,$\cite{Kaiser Leutwyler} concerning the definition of 
$\chi(q^2)$. The correlation function
$\lvac T\,\omega(x)\,\omega(0)\rvac$ is too singular for the
integral in eq.~(\ref{eq:ts}) to make sense as it stands.
The corresponding dispersion relation contains two
subtractions:$\,$\cite{Minkowski}
\bea \chi(q^2)=\chi(0) + q^2\chi'(0)+\frac{q^4\!}{\pi}\!\int
\frac{ds}{s^2(s-q^2-i\epsilon)}\,\mbox{Im}\,\chi (s)\fs\nonumber\eea 
The first one of these is fixed by the invariance of the 
effective action under local chiral rotations, but the second is
not:
(i) For the path integral
over the quarks and gluons to make sense 
in the presence of external fields, all terms of mass dimension
$\leq 4$ that are consistent with the symmetries of the theory must be
included in the QCD Lagrangian. (ii) The invariance property
(\ref{eq:anomalies}) excludes a term
proportional to $\theta^2$, but does allow one of the
form $h_0 D_\mu\theta D^\mu \theta$. In fact, such a term is needed
to absorb the quadratic divergences occurring in the perturbation theory 
graphs relevant
for the topopogical susceptibility. Hence the value of $\chi'(0)$ depends
on the method used when subtracting these infinities. How come that we
can make a statement about its numerical magnitude ?

The reason is that all of the graphs contributing to the correlation function 
either remain finite or disappear when $N_c$ is sent to infinity. 
The same is true of the
renormalization ambiguity contained therein: On the level of the effective
theory, the constant $h_0=O(1)$ exclusively contributes to $H_0=O(1)$. 
Since $\chi'(0)$ is a quantity of $O(N_c)$,
the problem only shows up at nonleading orders
of the $1/N_c$ expansion. The formula (\ref{eq:chiprim})
shows that the leading term in the simultaneous 
expansion of $\chi'(0)$ in
powers of $m_u,m_d,m_s$ and $1/N_c$ is fully determined by the quark mass
ratios and by the pion decay constant. In the evaluation of the sum rules,
the problem does not show up, because the perturbative contributions that 
would require renormalization are
discarded -- the results obtained concern the nonperturbative
contributions to $\chi'(0)$.

\section{Higher orders}

At next-to-leading order, the Langrangian involves further effective
coupling constants. The effective theory contains infinitely
many such constants, which chiral symmetry leaves undetermined  -- 
they represent the analogues of the Taylor coefficients
ocurring in the Heisenberg-Euler Lagrangian. There is a difference in that
those coefficients can explicitly be calculated in terms of the electron mass,
while an explicit expression for the effective
coupling constants $F,\,B,\,\ldots$ in terms of the scale of QCD 
is not available. Quite a few of these have been determined on the basis of
experimental information and for some, a numerical determination on the
lattice has been performed.$\,$\cite{LEC on lattice} 

As mentioned above, the one-loop graphs of ${\cal
  L}^{(2)}$ represent contributions of order $p^4$. Dimensional regularization
preserves the symmetries of the Lagrangian. The divergences arising from the
one-loop graphs thus represent local terms that are invariant under 
U(3)$_{\indR}\times$U(3)$_{\indL}$. Since the Lagrangian ${\cal L}^{(4)}$
contains all terms permitted by the symmetry, the divergences contained in the
one-loop graphs of ${\cal L}^{(2)}$ can be
absorbed in a renormalization of the effective coupling constants in
${\cal L}^{(4)}$. The argument extends to graphs with an arbitrary number
of loops, including those that involve vertices from the
higher order terms in the derivative expansion of the effective Lagrangian. 

Taken by itself, the nonlinear $\sigma$-model  
does not make sense, because the divergences generated by the
quantum fluctuations cannot be absorbed by renormalizing the coupling
constants $F$, $B$. Indeed, the model only represents the leading term in the 
derivative expansion of the full effective Lagrangian. The effective theory
does provide the proper embedding for the model to be meaningful: 
The divergences can be absorbed in the couplings
of higher order. The results obtained on the
basis of the effective theory to any given order in the low energy expansion 
are unambiguous. In particular, they do not depend on the regularization
used -- in this sense, the effective theory represents a renormalizable
framework that involves infinitely many coupling constants. 
  
Finally, I recall that the full effective action of QCD
is not invariant under chiral rotations, because of the anomalies. 
For the effective theory to
account for the anomalous terms in the Ward identities, the  effective 
Lagrangian must contain contributions that are not invariant. 
In fact, a closed expression for the relevant contributions is known since a 
long time: the Wess-Zumino-Witten Lagrangian.$\,$\cite{WZW} 
The full effective Lagrangian
is obtained by first writing down all possible vertices that are
invariant under U(3)$_{\indR}\times$U(3)$_{\indL}$ and then adding this 
term, which represents a contribution of $O(p^4)$ and thus belongs to
${\cal L}^{(4)}$. The WZW-term does not involve any free parameters -- like
the anomalies themselves, it represents a purely geometric contribution
that is fully determined by the number of colours.

\section{Outline of the proof}\label{proof}
I now briefly sketch the proof$\;$\cite{foundations} of the claim that (a) the
low energy expansion of the effective action of QCD can be worked out by means
of an effective field 
theory and (b) the relevant effective Lagrangian is invariant under local
chiral rotations, except for the WZW-term. 
The basic hypothesis underlying this proof is that the Goldstone bosons
required by spontaneous symmetry breakdown are the only massless particles
contained in the spectrum of asymptotic states, so that only these generate
singularities at low energies. 

One first shows that, as a consequence of the Ward identities, Goldstone
bosons of zero momentum cannot interact. This is crucial, because it implies
that the interaction becomes weak at low energies -- the reason why the low
energy properties of QCD can be worked out explicitly, despite the fact that
the interaction among the quarks and gluons is strong there. 

Next, one considers the Goldstone boson scattering amplitude. 
The structure of the low energy singularities contained therein
is determined by the cluster decomposition: One-particle exchange generates
poles, while the exchange of two or more particles produces branch cuts.
Since Goldstone bosons of zero momentum do not interact, the scattering
amplitude disappears if all of the momenta are sent to zero. Unitarity implies
that the imaginary part of the scattering amplitude is given by its square. In
four dimensions, the relevant phase space factors yield nonnegative powers of
the momentum, so that the imaginary part disappears more rapidly than the real
part when the momenta tend to zero: 
The contributions generated by the exchange of two or more particles only show
up at nonleading orders of the low energy expansion. The leading term
exclusively contains the poles due to one-particle exchange. At leading order,
their residues -- the one-particle irreducible parts -- are free of
singularities and can thus be expanded in the momenta. This property underlies
all of the early work on current algebra and PCAC and used to be referred to
as the pion pole dominance hypothesis. 

The one-particle irreducible parts vanish at zero momentum. Lorentz invariance
and Bose statistics thus imply that the leading term in their low energy
expansion is of the form 
\bea -\hspace{-1.5em}\sum_{\mbox{\footnotesize perm}\;(1,\,\ldots\,,\,n)}
\hspace{-1.5em}\left\{ g \rule[-0.3em]{0em}{0em}_{i_1\ldots i_n}
\;p_1\cdot p_2 + h \rule[-0.3em]{0em}{0em}_{i_1\ldots i_n}\;
p_1^2\right\}\nonumber\co\eea
where $p_1\,\ldots\,,\,p_n$ are the momenta flowing into the irreducible
part in question and $i_1,\ldots\,,\,i_n$ label the flavour 
quantum numbers of the corresponding Goldstone bosons. 
The generic form of the leading contributions 
in the low energy expansion of the scattering amplitude is given by a product
of pole terms and factors of the above form. Since contributions
proportional to the square of the momentum of one of the particles cancel
against the corresponding pole term, we may without loss of generality set $ h
\rule[-0.3em]{0em}{0em}_{i_1\ldots i_n}=0$.

At leading order of the low energy expansion, 
the scattering amplitude is of the same structure
as the tree graphs of a field theory. In fact, the tree graphs of the 
Lagrangian
\bea {\cal L}^{(2)}=
\sum_{i,\,k}g\rule[-0.3em]{0em}{0em}_{ik}(\pi)
\partial_\mu\pi^i\partial^\mu\pi^k \,,\hspace{1em}
g\rule[-0.3em]{0em}{0em}_{ik}(\pi)=\mbox{$\frac{1}{2}$}\,\delta
\rule[-0.3em]{0em}{0em}_{ik}+\sum_{n=3}^\infty\hspace{-0.8em} 
\sum_{\hspace{1em}i_3,\,\ldots\,,\,i_n}\!\!\!
g \rule[-0.3em]{0em}{0em}_{i k\hspace{0.04em} i_3\ldots
  i_n}\pi^{i_3}\ldots\pi^{i_n}\nonumber\eea
generate precisely the same scattering amplitude.
The analysis can be extended to all orders of the low energy expansion.
Clustering implies that the leading vertices also determine the
leading contributions to the low energy singularities generated by
multiparticle exchange. At next-to-leading order, only the cuts due to
two-particle exchange contribute. Removing these, the one-particle irreducible
parts are free of singularities up to and including $O(p^4)$.
The contributions of order $p^4$ may again be represented by
corresponding vertices in the effective Lagrangian,
etc. 

In order to extend the effective theory from the scattering amplitude
to the Green functions of
QCD, one needs to analyze the low energy expansion of the matrix elements
of the currents with the Goldstone bosons. The singularities occurring 
therein can be
sorted out in the same manner as for the scattering amplitude. The net result
is that it suffices to equip the effective Lagrangian with suitable vertices,
which in addition to the field $\pi^i(x)$ and its derivatives also involve
the external fields $v_\mu(x),\,a_\mu(x),\,m(x),\,\theta(x)$ and their
derivatives. This then establishes claim (a).

Up to here, the analysis is perfectly general and applies to any system with a
spontanously broken Lie group.  
If ${\cal L}_{\eff}$ is invariant under a local group of
symmetries, then the same holds for the effective action that it generates --
dimensional regularization manifestly preserves the gauge invariance of a
scalar field theory, so that the symmetries of the classical Lagrangian also
represent symmetries of the path integral. Claim (b) states that, for Lorentz 
invariant theories in four dimensions such as QCD, the converse is
also true: The symmetries of the effective action imply that the effective
Lagrangian can be brought to invariant form, except for the WZW term.
The statement does not hold in the general case. In three dimensions, for
instance, 
Chern-Simons theory represents a counter example. In four dimensions, the 
effective Lagrangian relevant for the spin waves of a ferromagnet does not
have the same symmetries as the corresponding effective action: Under local
spin rotations, the relevant Lagrangian changes by a total 
derivative.$\,$\cite{nonrelativistic} 

The main problem encountered in the analysis of the symmetry properties of the
effective theory is that the dynamical variables $\pi^i(x)$ do not have
immediate physical significance. They merely serve as variables of integration
in the path integral -- the underlying theory does not contain such
quantities. One may, for instance, subject the 
variables to a transformation of the type $\pi^{i}(x)'=
f^i(\pi)$ without changing the content of the effective theory. The freedom
corresponds to the fact that the off-shell extrapolation of matrix elements
such as $\lvac A^\mu_i|\pi^k\rangle$ or of the scattering amplitude 
is arbitrary. The explicit form of ${\cal L}_{\eff}$, however,
does depend on the choice of the variables.
This implies that the effective Lagrangian is not unique -- a circumstance
that makes it rather tedious to establish its properties. In practical
applications, the problem manifests itself through the freedom of adding
terms of nonleading order to the effective Lagrangian that are 
proportional to the classical
equation of motion. Since such terms can be removed with a suitable change
of variables, they are irrelevant. The various contributions generated by 
two Lagrangians that only differ in this 
manner, however, are not the same -- only the sum relevant 
for physical quantities is. 

At leading order of the low energy expansion, the effective action is given by
the tree graphs of ${\cal L}^{(2)}$, that is by the extremum of the
corresponding classical action. 
The tensor $g\rule[-0.3em]{0em}{0em}_{ik}(\pi)$, that collects the
effective coupling constants associated with the leading order 
Goldstone boson interaction vertices, plays the role of a metric
on the space of the effective fields. The invariance of the effective action
implies that this metric admits a group of isometries. The relevant Killing
vectors also show up in the effective Lagrangian: They represent the 
coefficients
of the terms that are linear in the external vector and axial fields.
The effective fields $\pi^i$ may be viewed as the coordinates of the quotient 
space G/H, where G and H are the symmetry groups of the Hamiltonian and of the
vacuum, respectively. This space carries an intrinsic metric: The one induced 
by the metric on the Lie group G. In the case of QCD, where G/H = SU(3), 
the metric relevant for the leading term in the derivative expansion of the
effective Lagrangian differs from the intrinsic metric of the group SU(3) 
only by an overall factor, given by the square of the pion decay constant: 
\bea ds^2 =\sum_{i,k}g \rule[-0.3em]{0em}{0em}_{i
  k}(\pi)\,d\pi^id\pi^k=\mbox{$\frac{1}{4}$}\,F^2 \langle dU
dU^\dagger\rangle\fs\nonumber\eea 
This shows that the low energy strucure of the theory is determined by group 
geometry and explains why a model of mathematical physics turns out
to be of relevance for our understanding of nature:
For symmetry reasons, the leading order effective Lagrangian of QCD is the one 
of the nonlinear $\sigma$-model. 

The analysis extends to all orders of the derivative
expansion:$\,$\cite{foundations} With a
suitable choice of the variables, the effective Lagrangian is invariant under
local chiral rotations, except for the contributions generated by the
anomalies. 

\section{Effective Lagrangian of next-to-leading order}
By construction, the effective theory yields the general solution of the Ward
identities obeyed by the Green functions of QCD. At leading order of the
low energy expansion, this solution is fully determined by the three
constants $F$, $B$ and $H_0$. At next-to-leading order, the general solution
of the Ward identities contains 12 additional parameters, even 
if the singlet vector and axial currents and the winding number density are
omitted. Their inclusion requires 11 further effective coupling constants --
the explicit expression for ${\cal L}^{(4)}$ contains a plethora
of terms. 

Lorentz invariance implies that the Green functions can be decomposed into 
scalar functions,
with coefficients that contain the external momenta and the tensors
$g_{\mu\nu}$, $\epsilon_{\mu\nu\rho\sigma}$. In view of the fact that the
square of $\epsilon_{\mu\nu\rho\sigma}$ can be expressed in terms of 
$g_{\mu\nu}$, there are 
two categories of contributions: The natural parity part of the effective
action, which collects the pieces that do not contain the $\epsilon$-tensor, 
and the unnatural parity part, where this
tensor occurs exactly once. 
Since the contributions from the anomalies
are proportional to the tensor $\epsilon_{\mu\nu\rho\sigma}$, 
they only affect the unnatural parity part of the effective
action. For the natural parity part, chiral symmetry thus amounts to a very 
simple statement: This part of the effective action is invariant under
local chiral rotations of the external fields.

It is convenient to decompose the effective Lagrangian accordingly.
The natural parity part, in particular, the leading term of the derivative
expansion, is invariant under local chiral rotations.
Most of the terms occurring at first nonleading order also belong to 
the natural parity part of the effective Lagrangian. The full expression for
this part reads:$\,$\cite{Kaiser Leutwyler} 
\bea {\cal L}^{(4)}_{np}\al=\al L_1\,\langle D_\mu U^\dagger D^\mu U\rangle^2 +
L_2\, \langle D_\mu U^\dagger D_\nu U\rangle \langle D^\mu U^\dagger D^\nu U
\rangle\no\al\al +L_3\, \langle D_\mu U^\dagger  D^\mu U
D_\nu U^\dagger D^\nu U\rangle+ L_4\, \langle D_\mu U^\dagger D^\mu U\rangle
\langle U^\dagger\chi  +  \chi^\dagger U\rangle\no\al\al+
L_5\, \langle D_\mu U^\dagger D^\mu U (U^\dagger \chi+  \chi^\dagger U)\rangle
+L_6\,\langle U^\dagger\chi  +\chi^\dagger U\rangle^2\no\al\al +L_7\,\langle 
U^\dagger \chi- \chi^\dagger U\rangle^2 + L_8\,\langle U^\dagger\chi
U^\dagger\chi +\chi^\dagger U \chi^\dagger U\rangle\\
\al\al-i L_9\,\langle F^{\indR}_{\mu\nu} D^\mu U D^\nu U^\dagger+
F^{\indL}_{\mu\nu} D^\mu U^\dagger D^\nu U\rangle+
L_{10}\, \langle F^{\indR}_{\mu\nu}U
F^{\indL\,\mu\nu} U^\dagger\rangle\no\al\al - i L_{11}\,\, D_\mu \theta\,\br
U^\dagger 
D^\mu U D_\nu U^\dagger D^\nu U \ke + L_{12}\, D_\mu \theta D^\mu \theta\, \br
D_\nu U^\dagger D^\nu U \ke \no\al\al + L_{13}\, D_\mu \theta D_\nu  \theta\,
\br D^\mu U^\dagger D^\nu U\ke+ L_{14}\,D_\mu \theta D^\mu \theta \,\br
U^\dagger\chi +\chi^\dagger U \ke \no\al\al  
- i L_{15}\, D_\mu \theta\, \br D^\mu U^\dagger \chi-D^\mu U \chi^\dagger \ke 
+i L_{16}\,\partial_\mu D^\mu \theta \, \br U^\dagger \chi - \chi^\dagger U 
\ke \no \al\al +H_1\,\langle F^{\indR}_{\mu\nu} F^{\indR \mu\nu}+
F^{\indL}_{\mu\nu} F^{\indL \mu\nu}\rangle+H_2\,\langle\chi^\dagger 
\chi\rangle\no
\al\al +H_3\,v^0_{\mu\nu}v^{0\,\mu\nu}+
H_4 \,a^0_{\mu\nu}a^{0\,\mu\nu}  + 
H_5\,( D_\mu \theta D^\mu \theta )^2 + H_6 ( \partial_\mu D^\mu \theta )^2 
\co\nonumber\eea
with $\chi(x)\equiv 2 B m(x)$. Some of the couplings involve
the field strengths of the external fields: The traceless 
matrices $F^{\indR}_{\mu\nu}$, $F^{\indL}_{\mu\nu}$ collect the octet
components of the field strengths belonging to the right- and left-handed 
fields
$r_\mu=v_\mu+a_\mu$, $l_\mu=v_\mu-a_\mu$, respectively, 
while $v^0_{\mu\nu}$,
$a^0_{\mu\nu}$ are the abelian field strengths of the singlets 
$v_\mu^0$, $a_\mu^0$.

The main term in the unnatural parity part is the Wess-Zumino-Witten 
Lagrangian, which
does not involve unknown coupling constants, but there is
one extra term, relevant for Green functions involving the operators $A^\mu_0$
or $\omega$:  
\bea {\cal L}^{(4)}_{up}={\cal L}_{\mbox{\tiny WZW}}+i L_{17}\, 
\epsilon^{\mu\nu\rho\sigma} D_\mu \theta\, \br
F^{\indR}_{\nu\rho} D_\sigma U U^\dagger -  
F^{\indL}_{\nu\rho} U^\dagger D_\sigma U \ke
\fs\eea
A detailed discussion of the structure of the Wess-Zumino-term in the
presence of the external singlet fields $v_\mu^0$, $a_\mu^0$ and $\theta$ 
is given in the paper quoted above.$\,$\cite{Kaiser Leutwyler}
 
The coupling constants $L_1$, $L_2$ and $L_3$ multiply
vertices that contain four or more Goldstone bosons; these couplings, for
instance, occur at first nonleading order in the low energy representation of
the scattering amplitude. $L_4$ and $L_5$ determine the expansion of the
decay constants to first order in the quark masses and $L_6$, $L_7$, $L_8$ 
are relevant for the corresponding expansion of $M_\pi$, $M_K$, $M_\eta$.
The coupling constant $L_9$ enters, for example,
in the low energy expansion of the vector form factor, while $L_{10}$ 
is relevant for the decay $\pi\rightarrow e\nu\gamma$. As the constants
$L_{11},\,\ldots\,,L_{17}$ multiply vertices that disappear if the external
singlet fields $\theta$ and $\langle a_\mu\rangle$ are switched off, they only
matter for Green functions formed with the singlet operators $A^\mu_0$,
$\omega$, such as the topological susceptibility.

The couplings $H_1,\,\ldots\,,,H_6$ represent contact terms. Some of these
are subject to a renormalization problem similar to the one occurring
in $H_0$. The QCD Lagrangian, for instance, contains a term of the form 
$h_2\langle m^\dagger m\rangle$, which is needed to absorb the quadratic
divergences occurring in the graphs for the 
correlation functions of the scalar and pseudoscalar
densities. As a consequence, the matrix elements
$\lvac \ubar u\rvac$, $\lvac \dbar d\rvac$ and $\lvac \sbar s\rvac$ contain
a term linear in the quark masses that is inherently ambiguous.
Within the effective theory, the problem concerns the value of $H_2$ and
shows up already at leading order
in the $1/N_c$ expansion. In principle, it should be possible to fix the
ambiguity with the behaviour of the quark condensate when the
quark masses become large: On physical grounds, the condensate should
tend to zero -- this can be the case only for one particular choice of the
constant $h_2$, so that the value of $H_2$ is fixed by
this condition. 

\section{Perturbation theory}
The path integral of the effective theory may be evaluated perturbatively.
The leading term of the perturbation series
is given by the tree graphs of ${\cal L}^{(2)}$. At first nonleading order, 
both the tree graphs of ${\cal L}^{(4)}$ and the one-loop graphs generated by 
${\cal L}^{(2)}$ contribute. At next-to-next-to leading order, the two-loop
graphs of ${\cal L}^{(2)}$ and the one-loop
graphs containing one vertex from ${\cal L}^{(4)}$ need also be taken into
account, together with the tree graphs of ${\cal L}^{(6)}$, etc. 

The perturbative evaluation of the path integral is based on the decomposition
${\cal L}_{\eff}={\cal L}_{kin}+{\cal L}_{int}$, where the kinetic part
is quadratic in the fields $\pi^i(x)$. 
Most of the preceding discussion concerns
the properties of QCD when the quark masses are turned off, where
${\cal L}_{kin}=\frac{1}{2}\partial \pi\partial \pi$, so that
the perturbation series involves massless scalar propagators. The effects
generated by the quark masses are, however, accounted for in the effective
Lagrangian -- the perturbation series may just
as well be worked out for nonzero quark masses. The position of the poles
contained in the tree graphs of ${\cal L}^{(2)}$ are determined by
the contributions quadratic in $\pi^i(x)$. Ignoring the isospin breaking
due to the difference between $m_u$ and $m_d$, the eigenvalues are given by
\bea \Mo^2_{\!\pi}=2\hat{m}B\co\hspace{2em}
\Mo_{\!K}^2=(\hat{m}+m_s)B\co\hspace{2em}
\Mo_{\!\eta}^2=\mbox{$\frac{2}{3}$}(\hat{m}+2m_s)B\co\eea
with $\hat{m}=\frac{1}{2}(m_u+m_d)$.
The tree graphs of ${\cal L}^{(4)}$ and the one-loop graphs of ${\cal
  L}^{(2)}$ generate corrections of order $m^2$. At first nonleading order,
the result for the pion mass, for instance reads
\bea \label{eq:Mpi4}
M_\pi^2\al=\al \Mo^2_{\!\pi}\left\{1-\frac{16 \hat{m}B}{F^2}(L_5-2L_8)-
\frac{16(2\hat{m}+m_s)B}{F^2}(L_4-2L_6)\right.\no\al\al
+\left.\,\frac{1}{2F^2}\,\Delta(0,\Mo_{\!\pi})+
\frac{1}{6F^2}\,
\Delta(0,\Mo_{\!\eta})\right\}+O(m^3)\fs\eea
The term $\Delta(0,M)$ stands for the scalar propagator at the origin,
\bea \Delta(0,M)=\frac{1}{(2\pi)^4}\!\int 
\!\!\frac{dk}{M^2-k^2-i\epsilon}\fs\eea
It arises from a tadpole graph that describes the propagation
of a pseudoscalar particle of mass $M$, emitted and absorbed at one and 
the same point of space-time. The integral diverges quadratically. In
dimensional regularization, it contains a pole at $d=4$: 
\bea \Delta(0,M)\al=\al 2M^2\lambda +\frac{1}{16\pi^2}M^2 
\ln\frac{M^2}{\mu^2}\co\no
\lambda\al=\al\frac{\mu^{d-4}}{(4\pi)^2}\left\{\frac{1}{d-4}-\frac{1}{2}
(\ln 4\pi  +\Gamma'(1)+1)\right\} \fs\eea
The divergence can be absorbed in a renormalization of the effective 
coupling constants:
\bea \label{eq:ren}L_n= L_n^r + \Gamma_n \lambda\fs\eea
The pion mass stays finite when $d\rightarrow 4$, provided the coefficients
$\Gamma_n$ obey $\Gamma_5-2\Gamma_8=\frac{1}{6}$, 
$\Gamma_4-2\Gamma_6=-\frac{1}{36}$. Actually, the renormalization coefficients
of all of the coupling constants occurring at first nonleading order are
known. Those for $L_1\,\ldots\,,\,L_{10},\, H_1,\,H_2$ were worked out long
ago.$\,$\cite{GL 1985} The additional couplings 
$L_{11}\,\ldots\,,\,L_{17},H_3,\ldots\,,\,H_6$, which are needed to analyze 
the Green functions of the singlet currents and of the winding number
density, do not pick up renormalization -- the coefficients $\Gamma_{11}$,
$\Gamma_{12}$, $\ldots$ all vanish.$\,$\cite{Kaiser Leutwyler} 

The formula (\ref{eq:Mpi4}) shows that the expansion of $M_\pi^2$ in
powers of the quark masses does not represent a straightforward Taylor series,
but contains logarithmic terms of the type 
$m^2\ln m$. These so-called chiral logarithms are characteristic of chiral
perturbation theory and occur in many of the results. Their coefficient
is determined by chiral symmetry, in terms of the pion decay constant. 
In the above expressions, the scale of the logarithms is the running scale 
$\mu$ of dimensional regularization. The renormalized coupling constants
$L^r_n$ also depend on this scale. In the results of physical interest, 
the running scale drops out. If we wish, we may write the
above formula for $M_\pi$ in the form
\bea M_\pi^2\al=\al \Mo^2_{\!\pi}\left\{1+\frac{\Mo^2_{\!\pi}}{32\pi^2 F^2}\,
\ln \frac{\Mo_{\!\pi}^2}{\Lambda_{\ind A}^2}-
\frac{\Mo^2_{\!\eta}}{96\pi^2 F^2}\,
\ln \frac{\Mo_{\!\eta}^2}{\Lambda_{\ind B}^2}\right\}+O(m^3)\fs\eea
The two scales $\Lambda_{\ind A}$, $\Lambda_{\ind B}$ are independent of
$\mu$ -- their values are determined by the coupling constants
$L_4,\,L_5,\,L_6,\,L_8$. 

\section{Illustration: Form factors}
As a further illustration, I consider the electromagnetic form factor of the
pion, 
\bea \langle \pi^+(p')| j^\mu|\pi^+(p)\rangle=(p^\mu +p^{\mu\prime})\,
f_{\pi^+}(t)\fs\eea
In this case, perturbation theory leads to the following representation:
\bea \label{eq:form factor}f_{\pi^+}(t)=1+\frac{t}{F^2}\,\left\{2\,L_9+
2\phi(t,M_\pi)+\phi(t,M_K)\right\}+O(p^4)\fs\eea
The
leading term of the expansion is trivial -- it represents the charge of the
particle, $f_{\pi^+}(0)=1$.  At first nonleading order, there are two types of
contributions:  (i) The term proportional to $L_9$, which comes from a tree
graph containing a vertex from ${\cal L}^{(4)}$; it is linear in the
momentum transfer $t$.  (ii) The functions $\phi(t,M_\pi)$ and $\phi(t,M_K)$
are generated by one-loop graphs, which exclusively involve vertices from
${\cal L}^{(2)}$; they are nontrivial functions of $t$, containing branch
cuts that start at $t=4M_\pi^2$ and $t=4M_K^2$.  In dispersive language, the
cuts are generated by $\pi\pi$ and $K\bar{K}$ intermediate states.

The function $\phi(t,M)$ may be expressed in
terms of the scalar loop integral formed with two propagators:
\bea J(p^2,M)=\frac{1}{(2\pi)^4}\!
\int\!\frac{dk}{(M^2-k^2-i\epsilon)(M^2-(k-p)^2-i\epsilon)}\co\eea
which is logarithmically divergent, so that the divergent part is momentum
independent: The difference
$\bar{J}(t,M)\equiv J(t,M)-J(0,M)$ approaches a finite limit when
$d\rightarrow 4$. The explicit expression reads:
\bea J(t,M)\al=\al \bar{J}(t,M) - 2\lambda
-\frac{1}{16\pi^2}\left\{\ln\frac{M^2}{\mu^2}+1\right\}\co\no
\bar{J}(t,M)\al=\al\frac{1}{16\pi^2}\left\{\sigma \ln
  \frac{\sigma-1}{\sigma+1}+2\right\}\co\hspace{2em}\sigma
=\left\{1-\frac{4M^2}{t}\right\}^{\frac{1}{2}}\fs\nonumber\eea
In this notation, the function $\phi(t,M)$ is given by
\bea
\phi(t,M)\al=\al\frac{1}{12}\left\{(t-4M^2)\bar{J}(t,M)-2\,\lambda\,
  t-\frac{t}{16\pi^2}\left(\ln \frac{M^2}{\mu^2}+\frac{1}{3}\right)\right\}
\fs\nonumber\eea
The resulting explicit representation of the form factor shows that the
divergence of the one-loop contributions is absorbed by a 
renormalization of the coupling constant $L_9$ according to
eq.~(\ref{eq:ren}), with $\Gamma_9=\frac{1}{4}$. 

The corresponding expression for the charge radius becomes
\bea \langle r^2\rangle^\pi_{\ind V} =
\frac{12 L_9^r}{F^2}-\frac{1}{32 \pi^2 F^2}\left\{
2\ln \frac{M_\pi^2}{\mu^2}+\ln \frac{M_K^2}{\mu^2}+3\right\}+O(m)\fs\eea
The formula involves the coupling constant $L_9$.
Since the effective Lagrangian is consistent with chiral symmetry for any 
value of the coupling constants, symmetry alone does not determine
the charge radius. It does, however, relate different
observables.  The slope of the $K^0_{e_3}$ form factor $f_+(t)$, for instance, 
is also fixed by $L_9$.  Conversely, the experimental value of this
slope,$\,$\cite{PDG} $\lambda_+ = 0.0300\pm 0.0016$,
can be used to first determine the magnitude of $L_9$ and then to
calculate the pion charge radius.  This gives $\langle r^2\rangle^{\pi}_{\ind
  V} = 0.42\,\mbox{fm}^2$, to be compared with the experimental
result obtained by scattering pions on atomic 
electrons:$\,$\cite{Amendolia} $0.439\pm0.008$ fm$^2$.

In the case of the neutral kaon, the analogous representation reads 
\be
f_{K^0} (t) = \frac{t}{F^2} \{ - \phi_\pi (t) + \phi_K (t) \} + O(t^2, t m).
\ee A term of order one does not occur here, because the charge vanishes.
There is no contribution from ${\cal L}^{(4)}$, either.  Chiral
perturbation theory thus provides a parameter free prediction in terms of the
one-loop integrals $\phi_\pi(t), \phi_K(t)$.  In particular, up to corrections
of $O(m)$, the slope of the
form factor is given by$\,$\cite{GL form factors} 
\be \langle r^2\rangle^{K^0}_{\ind V} = -
\frac{1}{16\pi^2F^2} \ln \frac{M_K}{M_\pi} = - 0.04\; \mbox{fm}^2\co \ee to be
compared with the experimental value$\,$\cite{Molzon} 
$- 0.054 \pm 0.026$ fm$^2$. In the meantime, similar
parameter free one-loop predictions have been discovered for quite a few other
observables.$\,$\cite{reviews} 

The above expression for the charge radius exhibits another
interesting feature, which is related to the fact that
the cloud of Goldstone bosons 
that surrouds the pion becomes long range if the quark masses are
sent to zero: The charge radius tends to infinity in that limit. 
The phenomenon also shows up in the behaviour of
the form factor in the massless theory, where the expansion in powers of
the momentum transfer contains a nonanalytic term,
\bea f_{\pi^+}(t)=1-\frac{1}{64\pi^2 F^2}\;t \,\ln\frac{(-t)}{\Lambda_{\ind
  C}^2}+O(t^2)\co\eea 
so that the first derivative explodes at $t=0$.
The scale of the logarithm occurring here is fixed by the coupling constant
$L_9$:
\bea L_9^r= \frac{1}{128\pi^2}\,\left\{\ln\frac{\Lambda_{\ind
        C}^2}{\mu^2}-\frac{5}{3}\right\}\fs\eea  

For the present review, the above sample calculations must suffice to
illustrate the nature of the results obtained at first nonleading
order of the chiral perturbation series. Plenty of such one-loop results
are reported in the literature.$\,$\cite{reviews} In quite a few cases, 
the series has been worked out to next-to-next-to leading 
order, where the two-loop graphs give rise to double chiral 
logarithms.$\,$\cite{ff1}$^-$\cite{twopoint} 
The explicit form of ${\cal L}^{(6)}$ is known,$\,$\cite{Lag6}
as well as the renormalization of the effective coupling constants 
occurring therein.$\,$\cite{ren twoloop} The renormalization group flow
of the effective theory was also examined, in particular
in view of infrared attractive fixed points.$\,$\cite{rengroup}

Among other things, the evaluation of the e.m.~form factor to two loops 
allows a determination of the pion charge radius that
is free of the model assumptions underlying the ``experimental'' value quoted 
above. The result of the model independent analysis reads:$\,$\cite{ff2} 
$\langle r^2\rangle^{\pi}_{\ind V} = 0.437\pm 0.016\,\mbox{fm}^2$.

\section{Magnitude of the coupling constants} \label{coupling} One of the main
problems encountered in the effective Lagrangian approach is the occurrence of
an entire fauna of effective coupling constants.  If these constants are
treated as totally arbitrary parameters, the predictive power of the method is
nil --- as a bare minimum, an estimate of their order of magnitude
is needed.

In principle, the effective coupling constants $F, B, L_1, L_2, \ldots$ are
calculable.  They do not depend on the light quark masses, but are determined
by the scale $\Lambda_{\QCD}$ and by the masses of the heavy quarks.  The
available, admittedly crude evaluations of $F$ and $B$ on the lattice
demonstrate that the calculation is even feasible in practice.  As discussed
above, the coupling constants $L_1, L_2, \ldots$ are renormalized by the
logarithmic divergences occurring in the one-loop graphs.  This property sheds
considerable light on the structure of the chiral expansion and provides a
rough estimate for the order of magnitude of the effective coupling 
constants.$\,$\cite{Georgi Soldate}  
The point is that the contributions generated by the
loop graphs are smaller than the leading (tree graph) contribution only for
momenta in the range $\mid p \mid \raisebox{-0.3em}{$\stackrel{<}{\sim}$}\,
\Lambda_\chi$, where \be\label{chiral scale} \Lambda_\chi \equiv 4 \pi F/
\sqrt{N_f} \end{equation} is the scale occurring in the coefficient of the
logarithmic divergence ($N_f$ is the number of light quark flavours).  This
indicates that the low energy expansion is an expansion in powers of
$(p/\Lambda_\chi)^2$, with coefficients of order one.  The argument
also applies to the expansion in powers of $m_u, m_d$ and $m_s$, indicating
that the relevant expansion parameter is given by $(M_\pi/ \Lambda_\chi)^2$
and $(M_K/\Lambda_\chi)^2$, respectively.

A more quantitative picture may be obtained along the following
lines.$\,$\cite{GL 1983}
Consider again the e.m.~form factor of the pion and compare the chiral
representation (\ref{eq:form factor}) with the dispersion relation 
\bdm f_{\pi^+}(t) =
\frac{1}{\pi} \int^{\infty}_{4M^2_\pi} \frac{dt'}{t'-t}\, \mbox{Im} f_{\pi^+}
(t')\fs \edm In this relation, the contributions $\phi_\pi, \phi_K$ from the
one-loop graphs of chiral perturbation theory correspond to $\pi \pi$ and $K
\bar{K}$ 
intermediate states.  To leading order in the low energy expansion, the
corresponding imaginary parts are slowly rising functions of $t$.  The most
prominent contribution on the right-hand side, however, stems from the region
of the $\rho$-resonance, which nearly saturates the integral:  The vector meson
dominance formula, $f_{\pi^+} (t) = (1-t/M_\rho^2)^{-1}$, which results if all
other contributions are dropped, provides a perfectly decent representation of
the form factor for small values of $t$.  In particular, this formula predicts
$\langle r^2\rangle^{\pi}_{\ind V} = 0.39$ fm$^2$, in satisfactory agreement
with observation.  This implies that the effective coupling
constant $L_9$ is approximately given by \be L_9 = \frac{F^2}{2M^2_\rho}\fs
\end{equation} In the channel under consideration, the pole due to $\rho$
exchange thus represents the dominating low energy singularity --- the $\pi
\pi$ and $K \bar{K}$ cuts merely generate a small correction.  More generally,
the validity of the vector meson dominance formula shows that, for the 
e.m.~form factor, the scale of the low energy expansion is set by 
$M_\rho = 770$ MeV.

Analogous estimates can be given for all effective coupling constants at order
$p^4$, saturating suitable dispersion relations with contributions from
reso\-nan\-ces,$\,$\cite{Ecker Gasser Pich de Rafael,Florida} for instance:  
\bdm L_5 =
\frac{F^2}{4M^2_S}\co\hspace{3em} L_7 = - \frac{F^2}{48M^2_{\eta'}}\co\edm
where $M_S \simeq 980$ MeV and $M_{\eta'} \simeq 958$ MeV are the masses of the
scalar octet and pseudoscalar singlet, respectively.  In all those cases,
where direct phenomenological information is available, these estimates do
remarkably well.  I conclude that the observed low energy structure is
dominated by the poles and cuts generated by the lightest particles --- hardly
a surprise.  

The effective theory is constructed on the asymptotic states of QCD.  In the
sector with zero baryon number, charm, beauty, $\ldots\,,$ the Goldstone
bosons form a complete set of such states, all other mesons being unstable
against decay into these (strictly speaking, the $\eta$ occurs among the
asymptotic states only for $m_d=m_u$; it must be included among the degrees of
freedom of the effective theory, never\-theless, because the masses of the 
light
quarks are treated as a perturbation --- in massless QCD, the poles generated
by the exchange of this particle occur at $p=0$).  The Goldstone degrees of
freedom are explicitly accounted for in the effective theory --- they
represent the dynamical variables.  All other levels manifest themselves only
indirectly, through the values of the effective coupling constants.  In
particular, low lying states such as the $\rho$ generate relatively small
energy denominators, giving rise to relatively large contributions to some of
these coupling constants.

In some channels, the scale of the chiral expansion is set by $M_\rho$, in
others by the masses of the scalar or pseudoscalar states occurring around
1 GeV.  This confirms the rough estimate (\ref{chiral scale}).  The cuts
generated by Goldstone pairs are significant in some cases and are negligible
in others, depending on the numerical value of the relevant Clebsch-Gordan
coefficient.  If this coefficient turns out to be large, the coupling constant
in question is sensitive to the renormalization scale used in the loop graphs.
The corresponding pole dominance formula is then somewhat fuzzy, because the
prediction depends on how the resonance is split from the continuum underneath
it. 

More precise results can be obtained by evaluating suitable dispersion
integrals or sum rules, using unitarity to determine the relevant imaginary
parts.$\,$\cite{sum rules 1,sum rules 2} This method makes it evident
that the pole dominance formulae only represent a crude parametrization: The
value of $M_S$, for instance, is the scale at which the relevant integral over
the imaginary part receives its main contributions -- it is inessential
whether or not that contribution is adequately described by a narrow
peak.$\,$\cite{Wang Yan} 

The quantitative estimates of the effective couplings given above explain why
it is justified to treat $m_s$ as a perturbation.  At order $p^4$, the
symmetry breaking part of the effective Lagrangian is determined by the
constants $L_4, \ldots, L_8$.  These constants are immune to the low energy
singularities generated by spin 1 resonances, but are affected by the exchange
of scalar or pseudoscalar particles, so that their magnitude is
determined by the scale $M_S \simeq M_{\eta'} \simeq 1$ GeV.  Accordingly, the
expansion in powers of $m_s$ is controlled by the parameter $(M_K/M_S)^2
\simeq \frac{1}{4}$.  The asymmetry in the decay constants, for example, is
determined by $L_5$.  The estimate of this coupling constant given above
yields \bdm \frac{F_K}{F_\pi} \simeq 1+\frac{M^2_K - M^2_\pi}{M^2_S}
\co \edm up to chiral logarithms and higher order terms.
This shows that
the breaking of the chiral and eightfold way symmetries is controlled by the
mass ratio of the Goldstone bosons to the non-Goldstone states of spin zero.
In chiral perturbation theory, the observation that the Goldstones are the
lightest hadrons thus 
acquires quantitative significance:  For momentum independent quantities such
as masses, decay constants, charge radii or scattering lengths, the magnitude
of consecutive orders in the chiral perturbation series is determined by 
the ratio $(M_K/M_S)^2$.

\section{Partition function}
Chiral perturbation theory yields remarkable insights into the equilibrium 
properties of the theory at temperatures below the chiral phase transition.
The same effective Lagrangian that provides a representation for the Green
functions of QCD also yields a representation for the
partition function,$\,$\cite{GL temperature} 
\bea \mbox{Tr} \exp \left(-\frac{H_{\QCD}}{kT}\right)={\cal N}_{\eff}^{\ind E}
\int\![d\pi] \exp \left(\mbox{$-\int\!d^3\!x\!\int_0^\beta \!dx^4$}\,
{\cal L}_{\eff}^{\ind E}\right)\co\eea
where ${\cal L}^{\ind E}_{\eff}$ is the euclidean form of the effective
Lagrangian, and the path integral extends over all configurations
that are periodic in the time direction, 
$\pi^i(\vec{x},x^4+\beta)=\pi^i(\vec{x},x^4)$, with $\beta=1/kT$. 
In particular, the melting of the quark condensate, which sets in when the
temperature rises,
can be worked out by means of this formula. If the masses $m_u$ and $m_d$ are
set equal to zero, the temperature expansion takes the 
form$,$\cite{Gerber Leutwyler}
\bdm\langle\ubar
u\rangle_{\hspace{-0.05em}\mbox{\raisebox{-0.1em} {\scriptsize $T$}}} =
\lvac\ubar u\rvac\!
\left\{1\,-\,\frac{T^2}
{8\bar{F}^2 }
\,-\,\frac{T^4}{384\bar{F}^4}
\,-\,\frac{T^6}{288\bar{F}^6}
\, \ln\left(\frac{T_1}{T}\right)
\,+\,O(T^8)\right\}\co\edm
where $\bar{F}$ is the value of $F_\pi$ in the limit $m_u=m_d=0$.
The formula is exact -- for massless quarks, the temperature scale relevant
at low $T$ is the pion decay constant. The additional logarithmic scale $T_1$
occurring at order $T^6$ is determined by the effective coupling constants that
enter the expression for the effective Lagrangian
at order $p^4$. Since these are known from the phenomenology of $\pi\pi$
scattering, the numerical value of $T_1$ is also known:$\,$\cite{Gerber
  Leutwyler}  $T_1=470\pm110\;\mbox{MeV}$. 

While the Goldstone bosons give rise to powers of
the temperature, massive states like the $\rho$-meson are suppressed by 
Boltzmann factors like $\exp(-M_\rho/kT)$. In view of the fact that
the spectrum contains many such states, these never\-theless generate a 
significant contribution, already for temperatures of order $140 \,\mbox{MeV}$
(there, the typical energy of the Goldstone bosons is of order 
$2.7\, k T\simeq 400\,\mbox{MeV}$). The massive states accelerate the melting
process. 

The low temperature expansion clearly exhibits the limitations of
the method: The truncated series can be trusted only at low temperatures,
where the first term represents the dominant contribution.  
In particular, the behaviour of
the quark condensate in the vicinity of the chiral phase transition is
beyond the reach of the effective theory discussed here.

The low temperature expansion was investigated for quite a few other 
quantities of physical interest. The position and the residue of the
pole in the thermal correlation function of the axial current
(effective values of $M_\pi$ and $F_\pi$) are known up to and including two 
loops.$\,$\cite{Schenk Toublan} For massive particles, the sensitivity of the
mass and of the width to the temperature was also analyzed 
in detail.$\,$\cite{temperature} These results are of interest, in particular, 
for the physics of the final state produced in heavy ion collisions.
Some of the corresponding transport coefficients have been worked out by
means of chiral perturbation theory.$\,$\cite{Goity}
The effects due to an external magnetic field$\,$\cite{Agasian} 
have also been investigated. Recently, the 
behaviour of QCD at high chemical potential, in particular, the
occurrence of a ``colour--flavour locking phase'' has attracted
great interest.$\,$\cite{Wilczek} The properties of such a phase can also be 
analyzed by means of a suitable effective Lagrangian.$\,$\cite{Gatto}

\section{Universality}
The Higgs sector of the Standard Model is also characterized by a 
spontaneously broken symmetry. In that case, the 
Hamiltonian is symmetric under G = O(4), while the symmetry group of the 
ground state
is the subgroup of those rotations that leave the expectation
value $\lvac\vec{\varphi}\rvac$ of the Higgs field invariant, 
H = O(3). As discussed in section
\ref{proof}, the structure of the effective Lagrangian follows from
the Ward identities obeyed by the Green functions. The form of these
identities is controlled by the structure of G and H in the infinitesimal
neighbourhood of 
the neutral element. Since the groups O(4) and O(3) are locally isomorphic to
SU(2)$_{\indR}\times$SU(2)$_{\indL}$
and SU(2)$_{\indV}$, respectively,
the effective Lagrangians relevant for the Higgs model and for
QCD with two massless flavours are identical. At the level of the
effective theory, the only difference between these two physically quite
distinct systems is that the numerical values of the effective coupling
constants are different. In the case of QCD, the one occurring at leading
order of the derivative expansion is the pion decay constant, $\bar{F}\simeq
90\,\mbox{MeV}$, while in the Higgs model, this coupling constant is larger
by more than three orders of magnitude, $\bar{F}\simeq
250\;\mbox{GeV}$. At next--to--leading order, the effective coupling constants
are also different. In particular, in QCD, the anomaly coefficient is equal to
$\mbox{N}_c$, while in the Higgs model, it vanishes.

The operators relevant for the expectation values $\lvac \qbar
q\rvac$, $\lvac\vec{\varphi}\rvac$ transform in the same manner under G.
The above formula for the quark condensate thus holds, without any change
whatsovever, also for the Higgs condensate:
\bdm \langle\vec{\varphi}\rangle_{\hspace{-0.05em}\mbox{\raisebox{-0.1em} 
{\scriptsize $T$}}} =
\lvac\vec{\varphi}\rvac\!
\left\{1\,-\,\frac{T^2}
{8\bar{F}^2 }
\,-\,\frac{T^4}{384\bar{F}^4}
\,-\,\frac{T^6}{288\bar{F}^6}
\, \ln\left(\frac{T_1}{T}\right)
\,+\,O(T^8)\right\}\fs\edm
In fact, the universal term of order $T^2$ was discovered in the framework of
the Higgs model, in connection with work on the electroweak phase
transition.$\,$\cite{Binetruy}

The example illustrates the physical nature of effective theories: At long
wavelength, the microscopic structure does not play any role. The behaviour
only depends on those degrees of freedom that require little
excitation
energy. The hidden symmetry, which is responsible for the absence of an
energy gap and for the occurrence of Goldstone bosons, at the same time also
determines their low energy properties. For this reason, the form of
the effective Lagrangian is controlled
by the symmetries of the system and is, therefore, universal.
The microscopic structure of the underlying theory exclusively manifests
itself in the numerical values of the effective coupling constants. 

Chiral perturbation theory may also be used to analyze the
spontaneous breakdown of electroweak gauge symmetry, without relying on the
assumption that the phenomenon is described by the Higgs model. As far as the
low energy structure is concerned, alternative models such as
technicolour are described by the same effective theory -- 
the degrees of freedom
involved in the formation of the electroweak condensate only show up 
indirectly, in the values of the effective coupling 
constants.$\,$\cite{electroweak,Nyffeler} 

\section{Concluding remarks}
The main motivation for working in chiral dynamics is that it is fun, but for
those strolling in other fields,  
one can give a few scientific reasons that indicate what this is good for.

In condensed matter physics, effective theories have successfully been used 
since a long time. In particular, the phenomenon of spontaneous symmetry
breakdown was discovered there. In this perspective, chiral perturbation
theory is just another illustration of the fact that the relevant degrees of
freedom must be identified to arrive at a good understanding of 
the physics. Lorentz invariance implies that the relation between the
wavelength and the frequency of the
Goldstone bosons of particle physics is given by 
$\omega=c|\vec{k}|$.  The dispersion law for the magnons of an antiferromagnet
is of the same form, but there are two differences: The value of $c$
is not the same and, more importantly, the relation is linear in $|\vec{k}|$
only for large wavelengths -- the dispersion law 
also contains higher powers of $\vec{k}$, while Lorentz invariance excludes 
such terms.  The effective theories
describing the magnons of a ferromagnet or the phonons of a
solid differ even more from those relevant for particle physics -- it is very
instructive to see why that is so.$\,$\cite{nonrelativistic}

Chiral perturbation theory has become an indispensable tool in the 
phenomenological analysis, because it provides a detailed understanding of 
the low energy properties of the strong interactions and puts early attempts 
in this direction -- the static model of the $\pi N$ interaction, just to name
one -- on firm mathematical grounds. In particular, much of what we know
about the weak interactions comes from kaon decays. To analyze the relevant
observables, the effects due to the strong interactions must be
accounted for.

Also, what we know about the pattern of light quark masses heavily relies on
the results obtained with chiral perturbation theory. Gradually, lattice
calculations start contributing to our knowledge in this domain, 
but further progress with light dynamical fermions is required before 
the numbers obtained with this technique can be taken at face value,
particularly for $m_u$ and $m_d$ -- it is 
difficult to 
numerically simulate the effects generated by the emission and absorption of
the Goldstone bosons. Even so, these calculations already now shed a
considerable amount of light on the low energy properties of QCD. 

In connection with lattice simulations, chiral perturbation theory is useful
as a tool to analyze the finite size effects. Since these are dominated by the
lightest particles, that is by the Goldstone bosons, they can be calculated
on the basis of the effective theory.
For quite a few observables, the calculation has been done.$\,$\cite{finite
  volume} Once lattice
evaluations with dynamical quarks reach realistically small quark masses,
these results should turn out to be very useful, because they allow one to 
correct the numerical results for the most important finite size effects,
so that volumes of modest size should suffice -- in the standard approach, 
where the infinite volume limit is performed
by brute force, very large volumes are required.$\,$\cite{Luescher} 

The volume-dependence of the 
partition function is relevant 
also for an understanding of the spectrum of the Dirac operator. 
As pointed out by Banks and Casher,$\,$\cite{Banks Casher} the quark 
condensate is determined by the spectral density at small eigenvalues. Chiral
perturbation theory allows one to establish a rather detailed picture
for the properties of the spectrum, as well as for the distribution of the 
winding number.$\,$\cite{Smilga} 

A further issue, where chiral perturbation theory turned out to
be very useful, is the Okubo-Iizuka-Zweig rule,
which becomes exact only in the limit where the number of colours 
is sent to infinity. The effective theory can be extended to cover
the case where $N_c$ is taken large. In this framework, the $\eta'$ plays a
crucial role, because the U(1)-anomaly, 
which is responsible for the bulk of the mass of this particle, 
is then suppressed.$\,$\cite{Witten Veneziano} The systematic
expansion of the effective theory in powers of $1/N_c$ allows one to
disentangle the effects that break the OZI rule from those that break flavour 
symmetry -- a necessary prerequisite to quantitatively describe, for instance,
$\eta-\eta'$ mixing.$\,$\cite{eta etaprim mixing} 
The extended effective theory 
also provides a handle on the ambiguity pointed out by 
Kaplan and Manohar,$\,$\cite{Kaplan Manohar} which is of relevance in 
connection with phenomenological determinations 
of the quark mass ratios: The ambiguity is suppressed to 
all orders of the $1/N_c$ expansion.$\,$\cite{Kaiser Leutwyler} 

One of the problems encountered in chiral perturbation theory is that
the truncated low energy expansion of the amplitudes satisfies unitarity 
only up to higher order contributions. In some cases, the effects generated by
the two-particle cuts are quite large, already at threshold -- in a one-loop
calculation, these are accounted for only to leading order of the low energy
expansion. In the case of the $\pi\pi$ scattering amplitude, a much more
accurate representation is obtained by matching the chiral representation
with a dispersive one, based on the Roy
equations.$\,$\cite{ACGL} Dispersive methods have also been used to analyze
the two-particle cuts in other amplitudes, such as two-point functions or
form factors.$\,$\cite{sum rules 1,sum rules 2,dispersive methods}   
An improved description may be obtained by rewriting the
chiral representation in such a manner that elastic unitarity is
obeyed exactly. In the case of a form factor, for instance, it suffices
to replace the one-loop formula for $f(t)$ by the one for $1/f(t)$. 
The Gounaris-Sakurai formula$\,$\cite{Gounaris Sakurai} 
amounts to precisely this prescription -- it provides a remarkably 
accurate low energy representation for the e.m.~form factor. As advocated by
Truong,$\,$\cite{Truong} a similar unitarization procedure can be applied to
other amplitudes as well. The main problem with this approach is that
the choice of the unitarization is not unique. It would be of considerable
interest to find out what reordering of the chiral perturbation series
is required to improve the representation in an algebraically controlled 
manner.
 
A fascinating aspect of chiral dynamics is the possibility
of subjecting the theory to experimental test. As an example, I 
briefly discuss the $S$-wave scattering lengths of $\pi\pi$ scattering. 
Since Goldstone bosons of zero momentum cannot interact, these scattering
lengths vanish if $m_u$ and $m_d$ are sent to zero -- like the pion mass,
they represent a quantitative measure for the breaking of 
chiral symmetry due to the quark masses.
In fact, in 1966, Weinberg showed that the two symmetry
breaking effects are related to one another. In particular, the leading order
prediction for  the isoscalar
scattering length reads:$\,$\cite{Weinberg pipi}
\bea a_0^0=\frac{7M_\pi^2}{32 \pi F_\pi^2}+O(m^2)\fs\nonumber\eea
The next-to-leading order terms in the low energy expansion of the 
$\pi\pi$ scattering amplitude were worked out$\,$\cite{GL 1983} in
1983 and, in 1996, those of next-to-next-to leading order 
were also calculated.$\,$\cite{BCEGS} Matching the two-loop representation of
the scattering amplitude with the dispersive representation obtained by solving
the Roy equations,$\,$\cite{ACGL,Wanders} the scattering lengths can be 
predicted
at the 2--3\% level of accuracy:$\,$\cite{CGL,Nieves Amoros} 
$a_0^0=0.220\pm 0.005$, 
to be compared with Weinberg's leading order formula, which yields
$a_0^0=0.16$ and with the one-loop result, $a_0^0=0.20$.
In this case, the convergence of the expansion in powers of the quark masses
is well understood -- the error bar attached to the above numerical 
result includes the uncertainties due to terms of yet higher 
order.

The example represents one of the very rare cases in strong interaction
physics, where theory is ahead of experiment. 
The symmetry breaking effects due to $m_u$, $m_d$ are hard to measure, because
they are very small. Currently, the best source of information concerning
the value of the $\pi\pi$ scattering lengths is the decay $K\rightarrow\pi\pi
e\nu$. The preliminary results of a recent measurement of this
decay$\,$\cite{Pislak} are consistent with the above prediction, but the
experimental errors still leave room for significant deviations. There is a 
beautiful proposal to due to Nemenov,$\,$\cite{DIRAC} based on the observation
that $\pi^+\pi^-$ atoms decay into a pair of neutral pions, through the
strong transition $\pi^+\pi^-\!\rightarrow\!\pi^0\pi^0$. Since the
momentum transfer nearly vanishes, the decay rate is proportional to
the square of the combination $a_0^0\!-\!a_0^2$ of $S$--wave $\pi\pi$ 
scattering lengths. The properties of the pionic atom have recently been 
analyzed at depth, on the basis of chiral perturbation theory,$\,$\cite{bound
  state} so that the precise form of the relation between the decay rate and 
the scattering lengths is known.
 Since the predictions for the latter are very sharp, the measurement of the
 lifetime of $\pi^+\pi^-$ atoms, which aims at an accuracy of 10\%,  
will provide 
a very stringent test of the standard framework that I have been relying on
throughout this review. This framework is based on the hypothesis that 
the Gell-Mann-Oakes-Renner relation is dominated by the 
order parameter of lowest dimension -- the quark condensate. 
As emphasized by J.~Stern and
collaborators,$\,$\cite{Stern Sazdjian Fuchs,sum rules 2} symmetry alone does 
not guarantee that this is so. If the outcome of the experiment should turn
out to be in conflict 
with the prediction quoted above, my understanding of chiral 
dynamics would undergo a first order phase transition.  

\section*{Acknowledgement} I thank Roland Kaiser for useful comments and
the Swiss National Science Foundation for support.


\section*{References}
\begin{thebibliography}{99}

\bibitem{Goldberger Treiman}
M.~Goldberger and S.~Treiman, {\it Phys.\ Rev.}~{\bf 110} (1958) 1178.

\bibitem{MENU99}For a detailed discussion, see 
Proc.\ MENU99,  Zuoz, Switzerland (1999), $\pi N${\it Newslett.}~{\bf
  15} (1999). 

\bibitem{Nambu} Y.~Nambu, {\it Phys.\ Rev.\ Lett.}~{\bf 4} (1960) 380.

\bibitem{Goldstone} J.~Goldstone, {\it Nuovo Cim.}~{\bf 19} (1961) 154.

\bibitem{Heisenberg}W.~Heisenberg, {\it Z.\ Phys.}~{\bf 77} (1932) 1.

\bibitem{Gell-Mann}M.~Gell-Mann and Y.~Ne'eman, {\it The Eightfold Way},\\ 
W.~A.~Benjamin (New York, 1964);\\
M.~Gell-Mann, {\it Physics} {\bf 1} (1964) 63.

\bibitem{Adler Weisberger}S.~L.~Adler, {\it Phys.\ Rev.}~{\bf 140} (1965)
  B736.\\
W.~I.~Weisberger, {\it Phys.~Rev.}~{\bf 143} (1966) 1302.

\bibitem{Weinberg multipi}S.~Weinberg, {\it Phys.~Rev.~Lett.}~{\bf 16} 
(1966) 879.

\bibitem{Weinberg pipi}S.~Weinberg, {\it Phys.~Rev.~Lett.}~{\bf 17} (1966)
616.

\bibitem{Callan} S.~Coleman, J.~Wess and B.~Zumino, {\it Phys.~Rev.}~{\bf 177}
  (1969) 2239;\\ C.~Callan, S.~Coleman, J.~Wess and B.~Zumino,\\  
{\it Phys.~Rev.}~{\bf 177} (1969) 2247.

\bibitem{Dashen Weinstein Li Pagels}R.~Dashen and M.~Weinstein, {\it
    Phys.~Rev.}~{\bf 183} (1969) 1291;\\
L.-F.~Li and H.~Pagels, {\it Phys.~Rev.~Lett.}~{\bf 26} (1971) 1204.

\bibitem{Weinberg 1979}S.~Weinberg, {\it Physica} A {96} (1979) 327.

\bibitem{GL 1983}J.\ Gasser and H.\ Leutwyler, {\it Phys.~Lett.}~B {\bf 125} 
(1983) 321;\\
{\it Ann.~Phys.~(N.Y.)} {\bf 158} (1984) 142.

\bibitem{GL 1985} J.\ Gasser and H.\ Leutwyler, {\it Nucl.~Phys.~}B {\bf 250}
(1985) 465.

\bibitem{reviews}
G.~Ecker,  in
{\it Quantitative Particle Physics}, Carg\`{e}se 1992, 
eds.~M.~L\'{e}vy {\it et al.}~(Plenum, New York, 1993) and in
{\it Broken Symmetries}, Schladming,
Austria (1998), hep-ph/9805500;\\
U.~Meissner, {\it Rep.~Prog.~Phys.~}{\bf 56} (1993) 903;\\
A.~Dobado, A.~Gomez-Nicola, J.~P.~Maroto and J.~P.~Pelaez,
{\it Effective Lagrangians for the Standard Model, Springer-Verlag, 
N.Y.~1997};\\
A.~Pich, in {\it Probing the Standard Model of Particle Interactions}, 
Les Houches, France (1997), 
hep-ph/9806303;\\
J.~Gasser,
%``Chiral perturbation theory,''
{\it Nucl.\ Phys.\ Proc.\ Suppl.}~{\bf 86} (2000) 257;\\
G.~Colangelo, hep-ph/0001256;\\
Barry R.~Holstein; hep-ph/0001281;\\
Jose Wudka, hep-ph/0002180.

\bibitem{Vafa Witten} 
C.~Vafa and E.~Witten,
%``Restrictions On Symmetry Breaking In Vector - Like Gauge Theories,''
{\it Nucl.~Phys.}~B {\bf 234} (1984) 173.

\bibitem{Gell-Mann Oakes Renner}M.~Gell-Mann, R.~J.~Oakes and B.~Renner, {\it
    Phys.~Rev.}~{\bf 175} (1968) 2195.

\bibitem{Heisenberg Euler}W.~Heisenberg and H.~Euler, {\it Z.~Phys.}~{\bf 98}
  (1936) 714.

\bibitem{foundations}H.~Leutwyler, {\it Ann.~Phys.~(N.Y.)} {\bf 235} (1994) 
165.

\bibitem{Crewther}R.~J.~Crewther, {\it Phys.~Lett.}~B {\bf 70} (1977) 349.

\bibitem{Di Vecchia Veneziano}P.~Di Vecchia and G.~Veneziano, {\it
    Nucl.~Phys.}~B {\bf 171} (1980) 253. 

\bibitem{Shore Veneziano}
G.~M.~Shore and G.~Veneziano,
%``The U(1) Goldberger-Treiman Relation And The Two Components Of The Proton
%'Spin','' 
{\it Phys.\ Lett.}~B  {\bf 244} (1990) 75;\\
S.~Narison, G.~M.~Shore and G.~Veneziano,
%``Target independence of the EMC - SMC effect,''
{\it Nucl.\ Phys.}~B  {\bf 433} (1995) 209;
%``Topological charge screening and the 'proton spin' beyond the chiral
%limit,'' 
{\it ibid.}~B {\bf 546} (1999) 235.

\bibitem{Ioffe}
B.~L.~Ioffe and A.~G.~Oganesian,
%``Proton spin content and {QCD} topological susceptibility,''
{\it Phys.\ Rev.}~D {\bf 57} (1998) 6590;\\
B.~L.~Ioffe,
%``Correlation of topological-charge densities at low Q**2 in QCD,''
{\it Phys.\ Atom.\ Nucl.}~{\bf 62} (1999) 2052;\\
%``Correlator of topological charge densities at low q**2 in {QCD}:
%Connection with proton spin problem,'' 
{\it Heavy Ion Phys.}~{\bf 9} (1999) 29;\\
B.~L.~Ioffe and A.~V.~Samsonov,
%``Correlator of topological charge densities in instanton model in {QCD},''
hep-ph/9906285.

\bibitem{quark mass ratios}H.~Leutwyler, {\it Phys.~Lett.}~B {\bf 378} (1996)
  313.

\bibitem{Kaiser Leutwyler}
R.~Kaiser and H.~Leutwyler, hep-ph/0007101.

\bibitem{Minkowski}
P.~Minkowski,
%``The Eta-Prime Meson Disobeys The Zweig Rule Up To Infinite Color,''
{\it Phys.\ Lett.}~B  {\bf 237} (1990) 531.

\bibitem{LEC on lattice}For a recent paper that also discusses earlier work,
  see\\
Jochen Heitger {\it et al.} [ALPHA collaboration], hep-lat/0006026.
 
\bibitem{WZW}J.~Wess and B.~Zumino,
%``Consequences Of Anomalous Ward Identities,''
{\it Phys.\ Lett.}~{\bf B37} (1971) 95;\\
E.~Witten,
%``Global Aspects Of Current Algebra,''
{\it Nucl.\ Phys.}~{\bf B223} (1983) 422.

\bibitem{nonrelativistic}H.~Leutwyler, {\it Phys.~Rev.}~D {\bf 49} (1994)
  3033;\\ {\it Helv.~Phys.~Acta} {\bf 70} (1997) 275;\\
C.~Hofmann, hep-ph/9706418; cond-mat/9805277;\\
Don H.~Kim and Patrick A.~Lee, {\it Ann.~Phys.}~{\bf 272} (1999) 130.

\bibitem{PDG} C.~Caso {\it et al.} [Particle Data Group], 
%``Review of particle physics,''
{\it Eur.\ Phys.\ J.}~{\bf C3} (1998) 1.

\bibitem{Amendolia} 
S.~R.~Amendolia {\it et al.} [NA7 collaboration],\\ 
{\it Nucl.~Phys.}~B {\bf 277} (1986) 168. 
 
\bibitem{GL form factors}J.\ Gasser and H.\ Leutwyler, 
{\it Nucl.~Phys.}~B {\bf 250} (1985) 517.

\bibitem{Molzon}W.~R.~Molzon, {\it Phys.~Rev.~Lett.}~{\bf 41} (1978) 1213.

\bibitem{ff1} J.~Gasser and U.~Meissner, {\it Nucl.~Phys.}~B {\bf 357} 
(1991) 90;\\
G.~Colangelo, M.~Finkemeier and R.~Urech,
{\it Phys.~Rev.}~D {\bf 54} (1996) 4403;\\
P.~Post and K.~Schilcher, hep-ph/0007095.

\bibitem{ff2}J.~Bijnens, G.~Colangelo and P.~Talavera,
%``The vector and scalar form factors of the pion to two loops,''
{\it JHEP} {\bf 9805} (1998) 014.

\bibitem{BCEGS}
J.~Bijnens, G.~Colangelo, G.~Ecker, J.~Gasser and M.~E.~Sainio,\\
%``Elastic $\pi\pi$ scattering to two loops,''
{\it Phys.\ Lett.}~B {\bf 374} (1996) 210;
%``Pion pion scattering at low energy,''
{\it Nucl.\ Phys.}~B {\bf 508} (1997) 263;\\
{\it ibid.}~{\bf B517} (1998) 639 (E).

\bibitem{twopoint}
E.~Golowich and J.~Kambor,
%``Chiral sum rules to second order in quark mass,''
{\it Phys.\ Rev.\ Lett.}~{\bf 79} (1997) 4092;\\
%``The DMO sum rule revisited,''
{\it Phys.\ Lett.}~B  {\bf 421} (1998) 319;
%``Two-loop analysis of axialvector current propagators in chiral
%perturbation theory,'' 
{\it Phys.\ Rev.}~D {\bf 58} (1998) 036004;\\
S.~Durr and J.~Kambor,
%``Two-point function of strangeness-carrying vector-currents in two-loop
%chiral perturbation theory,'' 
{\it Phys.\ Rev.}~D  {\bf 61} (2000) 114025;\\
G.~Amoros, J.~Bijnens and P.~Talavera,
%``Two-point functions at two loops in three flavour chiral perturbation
%theory,'' 
{\it Nucl.\ Phys.}~B {\bf 568} (2000) 319.

\bibitem{Lag6}H.~Fearing and S.~Scherer, 
%``Extension of the chiral perturbation theory meson Lagrangian to order
%p(6),'' 
{\it Phys.~Rev.}~D {\bf 53}
  (1996) 315;\\
J.~Bijnens, G.~Colangelo and G.~Ecker,
%``The mesonic chiral Lagrangian of order p**6,''
{\it JHEP} {\bf 9902} (1999) 020.

\bibitem{ren twoloop}J.~Bijnens, G.~Colangelo and G.~Ecker,
%``Double chiral logs,''
{\it Phys.\ Lett.}~B  {\bf 441} (1998) 437;
%``Renormalization of chiral perturbation theory to order p**6,''
{\it Ann.~Phys.~(N.Y.)}  {\bf 280} (2000) 100.

\bibitem{rengroup}D.~U.~Jungnickel and C.~Wetterich,
%``The linear meson model and chiral perturbation theory,''
{\it Eur.\ Phys.\ J.}~C  {\bf 2} (1998) 557;\\
M.~Atance and B.~Schrempp,
%``Infrared fixed points for ratios of couplings in the chiral Lagrangian,''
hep-ph/9912335.

\bibitem{Georgi Soldate} H.~Georgi and A.~Manohar, 
{\it Nucl.~Phys.}~B {\bf 234} (1984) 189; \\ 
M.~Soldate and R.~Sundrum, {\it Nucl.~Phys.}~B
{\bf 340} (1990) 1; 
 
R.~S.~Chivukula, M.~J.~Dugan and M.~Golden,\\ {\it Phys.~Rev.}~D
{\bf 47} (1993) 2930.

\bibitem{Ecker Gasser Pich de Rafael} G.~Ecker {\it et al., Nucl.~Phys.}~B
  {\bf 321} (1989) {311};\\ 
{\it Phys.~Lett.}~B {\bf 223} (1989) 425.

\bibitem{Florida} H.~Leutwyler, {\it Nucl.~Phys.}~B {\bf 337} (1990) 108 and
in {\it Yukawa couplings and the origin of
mass}, Proc.~2nd IFT
Workshop, University of Florida, Gainesville, Feb.~1994, ed.~P.~Ramond
(International Press, Cambridge MA, 1995).

\bibitem{sum rules 1}
J.~F.~Donoghue, J.~Gasser and H.~Leutwyler, \\ {\it Nucl.~Phys.}~B {\bf 343} 
(1990) 341.

\bibitem{sum rules 2}
B.~Moussallam,
%``N(f) dependence of the quark condensate from a chiral sum rule,''
{\it Eur.\ Phys.\ J.}~C {\bf 14} (2000) 111;
%``Flavor stability of the chiral vacuum and scalar meson dynamics,''
hep-ph/0005245;\\
S.~Descotes, L.~Girlanda and J.~Stern,
%``Paramagnetic effect of light quark loops on chiral symmetry breaking,''
{\it JHEP} {\bf 0001} (2000) 041;\\
S.~Descotes and J.~Stern, {\it Phys.~Rev.}~D {\bf 62} (2000) 504011;\\
%``Vacuum fluctuations of anti-q q and values of low-energy constants,''
hep-ph/0007082.

\bibitem{Wang Yan}X.-J.~Wang and M.-L.~Yan, hep-ph/0001150, hep-ph/0004157.

\bibitem{GL temperature}J.~Gasser and H.~Leutwyler, {\it Phys.~Lett.}~B {\bf
    184} (1987) 83.

\bibitem{Gerber Leutwyler}
P.\ Gerber and H.\ Leutwyler, {\it Nucl.~Phys.}~B {\bf 321} (1989) 387;\\
T.~Hatsuda,
%``Theoretical overview: Hot and dense QCD in equilibrium,''
{\it Nucl.\ Phys.}~A  {\bf 544} (1992) 27;\\
M.~C.~Birse,
%``Chiral symmetry in matter,''
Acta Phys.\ Polon.\  {\bf B29} (1998) 2357;\\
N.~O.~Agasian, D.~Ebert and E.~M.~Ilgenfritz,\\
%``Modelling the QCD phase transition with an effective Lagrangian of  light
% and massive hadrons,'' 
{\it Nucl.\ Phys.}~A {\bf 637} (1998) 135;\\
A.~V.~Smilga, %``Physics of hot and dense {QCD},''
{\it Nucl.\ Phys.}~A {\bf 654} (1999) 136C;\\
J.~R.~Pelaez,
%``The hadronic gas chiral phase transition within generalized chiral
% perturbation theory,'' 
{\it Phys.\ Rev.}~D  {\bf 59} (1999) 014002;\\
A.~Dobado and J.~R.~Pelaez,
%``Chiral symmetry and the pion gas virial expansion,''
{\it Phys.\ Rev.}~D  {\bf 59} (1999) 034004.

\bibitem{Schenk Toublan}
A.~Schenk, {\it Nucl.~Phys.}~B {\bf 363} (1991) 97;\\
D.~Toublan,
%``Pion dynamics at finite temperature,''
{\it Phys.~Rev.}~D {\bf 56} (1997) 5629.

\bibitem{temperature}H.~Leutwyler and A.~Smilga, {\it Nucl.~Phys.}~B
  {\bf 342} (1990) 302;\\
V.~L.~Eletsky and B.~L.~Ioffe,
%``Meson masses in nuclear matter,''
{\it Phys.\ Rev.\ Lett.}~{\bf 78} (1997) 1010;\\
%``Thermal mass shift of nucleons,''
{\it Phys.~Lett.}~B  {\bf 401} (1997) 327;\\
V.~L.~Eletsky, B.~L.~Ioffe and J.~I.~Kapusta,\\
%``Mass shift and width broadening of rho mesons produced in heavy ion
%  collisions,''
{\it Eur.\ Phys.\ J.}~A {\bf 3} (1998) 381;\\
S.~Mallik and K.~Mukherjee,
%``Rho parameters from odd and even chirality, thermal QCD sum rules,''
{\it Phys.\ Rev.}~D {\bf 61} (2000) 116007;\\
%``Thermal {QCD} sum rules with fewer power corrections,''
hep-ph/9809231;\\
V.~Sheel, H.~Mishra and J.~C.~Parikh,
%``Meson correlators at finite temperature,''
{\it Phys.\ Rev.}~D {\bf 59} (1999) 034501;\\
R.~Escribano, F.~S.~Ling and M.~H.~Tytgat,\\
%``Large N(c) chiral approach to M(eta') at finite temperature,''
{\it Phys.\ Rev.}~D  {\bf 62} (2000) 056004.

\bibitem{Goity}J.~L.~Goity and H.~Leutwyler, {\it Phys.~Lett.}~B {\bf 228}
  (1989) 517;\\
G.~M.~Welke, R.~Venugopalan and M.~Prakash,\\ {\it Phys.~Lett.}~B 
{\bf 245} (1990) 137;\\
H.~Bebie {\it et al., Nucl.~Phys.} ~B {\bf 378} (1992) 95;\\
J.~M.~Martinez Resco and M.~A.~Valle Basagoiti,\\
%``The speed of cool soft pions,''
{\it Phys.\ Rev.}~D  {\bf 58} (1998) 097901;\\
A.~Gomez Nicola and V.~Galan-Gonzalez,
%``Nonequilibrium chiral perturbation theory and pion decay functions,''
{\it Phys.\ Lett.}~B  {\bf 449} (1999) 288.

\bibitem{Agasian}N.~O.~Agasian and I.~A.~Shushpanov,
%``Quark and gluon condensates in a magnetic field,''
{\it JETP Lett.}~{\bf 70} (1999) 717;\\
%``The quark and gluon condensates and low-energy QCD theorems in a  magnetic
%field,'' 
{\it Phys.\ Lett.}~B  {\bf 472} (2000) 143;\\
N.~O.~Agasian,
%``Phase structure of the QCD vacuum in a magnetic field at low  temperature,''
hep-ph/0005300.

\bibitem{Wilczek}For a review, see\\
F.~Wilczek, hep-ph/9908480;\\
K.~Rajagopal, hep-ph/9908360.

\bibitem{Gatto}R.~Casalbuoni and R.~Gatto,
%``Effective theory for color-flavor locking in high density QCD,''
{\it Phys.\ Lett.}~B {\bf 464} (1999) 111;\\
%``The color-flavor locking phase at T not = 0: Exact results at order  T**2,''
{\it ibid.}~B {\bf 469} (1999) 213;
%``Pseudoscalar masses in the effective theory for color-flavor locking in
%high density QCD,'' 
hep-ph/9911223.

\bibitem{Binetruy}P.\ Bin\'{e}truy and M.\ K.\ Gaillard, 
{\it Phys.~Rev.}~D {\bf 32} (1985) 931.

\bibitem{electroweak}T.~Appelquist and C.~Bernard, {\it Phys.~Rev.}~D {\bf 22}
  (1980) 200;\\
A.~C.~Longhitano, {\it Phys.~Rev.}~D {\bf 22} (1980) 1166.

\bibitem{Nyffeler}
A.~Nyffeler and A.~Schenk,
%``Gauge-invariant Green's functions for the bosonic sector of the standard
% model,''
{\it Phys.\ Rev.}~D  {\bf 62} (2000) 036002;\\ 
A.~Nyffeler, %``The electroweak chiral Lagrangian revisited,'' 
hep-ph/9912472.
   
\bibitem{finite volume} J.~Gasser and H.~Leutwyler, {\it Phys.~Lett.}~B {\bf
  188} (1987) 477;\\
P.~Hasenfratz and H.~Leutwyler, {\it Nucl.~Phys.}~B {\bf 343} (1990) 241;\\
F.~C.~Hansen and H.~Leutwyler, {\it Nucl.~Phys.}~B {\bf 350} (1991) 201.\\
Chiral perturbation theory can also be applied to the quenched 
approxi\-mation on the lattice, see for instance:\\
C.~Bernard and M.~Golterman, {\it Phys.~Rev.}~D {\bf 46} (1992) 853;\\
S.~R.~Sharpe, {\it Phys.~Rev.}~D {\bf 46} (1992) 3146;
%``Enhanced chiral logarithms in partially quenched QCD,''
{\it Phys.\ Rev.}~D  {\bf 56} (1997) 7052;\\
M.~Golterman and K.~C.~Leung, {\it Phys.~Rev.}~D {\bf 56} (1997) 2950;\\
%``Applications of partially quenched chiral perturbation theory,''
{\it Nucl.\ Phys.\ Proc.\ Suppl.}~{\bf 73} (1999) 246;\\
G.~Colangelo and E.~Pallante, 
%``Quenched chiral perturbation theory to one loop,' 
{\it Nucl.~Phys.}~B {\bf 520} (1998) 433;\\
W.~Bardeen, A.~Duncan, E.~Eichten and H.~Thacker,
%``Anomalous chiral behavior in quenched lattice QCD,''
hep-lat/0007010.

\bibitem{Luescher}For an alternative approach, see\\ 
L.~Lellouch and M.~L\"uscher, hep-lat/0003023, and the references therein.

\bibitem{Banks Casher}T.~Banks and A.~Casher, {\it Nucl.~Phys.}~B {\bf 169}
  (1980) 103;\\
E.~Marinari, G.~Parisi and C.~Rebbi, {\it Phys.~Rev.~Lett.}~{\bf 47} (1981)
1795. 

\bibitem{Smilga}H.~Leutwyler and A.~Smilga, {\it Phys.~Rev.}~D {\bf 46} 
(1992)  5607;\\ 
A.~Smilga and J.~Stern,
%``On the spectral density of Euclidean Dirac operator in QCD,''
{\it Phys.\ Lett.}~B {\bf 318} (1993) 531;\\
J.~C.~Osborn, D.~Toublan and J.~J.~Verbaarschot,\\
%``From chiral random matrix theory to chiral perturbation theory,''
{\it Nucl.\ Phys.}~B {\bf 540} (1999) 317;\\
D.~Toublan and J.~J.~Verbaarschot,
%``The spectral density of the {QCD} Dirac operator and patterns of chiral  
% symmetry breaking,''
{\it Nucl.\ Phys.}~B {\bf 560} (1999) 259;
K.~Zyablyuk, {\it JHEP} {\bf 0006} (2000) 025;\\
Taro Nagao and Shinsuke M.~Nishigaki, hep-th/0001137;\\

\newpage
J.~B.~Kogut, M.~A.~Stephanov and D.~Toublan,\\
%``On two-color QCD with baryon chemical potential,''
{\it Phys.\ Lett.}~B {\bf 464} (1999) 183;\\
J.~B.~Kogut, M.~A.~Stephanov, D.~Toublan, J.~J.~Verbaarschot and A.~Zhitnitsky,
%``QCD-like theories at finite baryon density,''
{\it Nucl.\ Phys.}~B  {\bf 582} (2000) 477.

\bibitem{Witten Veneziano} 
E.~Witten, {\it Nucl.~Phys.}~B {\bf 156}
  (1979) 269;\\
G.~Veneziano, {\it Nucl.~Phys.}~B {\bf 159} (1979) 213.

\bibitem{eta etaprim mixing}R.~Kaiser and H.~Leutwyler, in
{\it Nonperturbative Methods in Quantum Field Theory}, eds. 
A.~W.~Schreiber, A.~G.~Williams and A.~W.~Thomas (World Scientific, 
Singapore, 1998);\\
E.~P.~Venugopal and B.~R.~Holstein,
%``Chiral anomaly and eta eta' mixing,''
{\it Phys.\ Rev.}~D  {\bf 57} (1998) 4397;\\
T.~Feldmann, P.~Kroll and B.~Stech,
%``Mixing and decay constants of pseudoscalar mesons,''
{\it Phys.\ Rev.}~D  {\bf 58} (1998) 114006;
%``Mixing and decay constants of pseudoscalar mesons: The sequel,''
{\it Phys.\ Lett.}~B {\bf 449} (1999) 339;\\
B.~Bagchi, P.~Bhattacharyya, S.~Sen and J.~Chakrabarti,\\
%``Mixing angles and decay constants of eta, eta-prime and eta(c),''
{\it Phys.\ Rev.}~D {\bf 60} (1999) 074002;\\
F.~Cao and A.~I.~Signal,
%``Two analytical constraints on the eta eta' mixing,''
{\it Phys.\ Rev.}~D  {\bf 60} (1999) 114012;\\
L.~S.~Celenza, B.~Huang and C.~M.~Shakin,
%``Importance of pseudoscalar-axial-vector mixing in calculation of  the
%properties of the pi, eta, and eta' mesons,'' 
{\it Phys.\ Rev.}~C {\bf 59} (1999) 2814;\\
R.~Escribano and J.~M.~Frere,
%``Phenomenological evidence for the energy dependence of the eta eta'  mixing
%angle,'' 
{\it Phys.\ Lett.}~B  {\bf 459} (1999) 288;\\
T.~Feldmann,
%``Quark structure of pseudoscalar mesons,''
{\it Int.\ J.\ Mod.\ Phys.}~A  {\bf 15} (2000) 159;\\
B.~Bagchi and P.~Bhattacharyya,
%``Determining the parameters of eta eta' mixing in the framework of
%anomalous Ward identities,'' 
{\it Mod.\ Phys.\ Lett.}~A  {\bf 15} (2000) 167.
 
\bibitem{Kaplan Manohar}D.\ B.\ Kaplan and A.\ V.\ Manohar, 
{\it Phys.~Rev.~Lett.}~{\bf 56} (1986) 2004.

\bibitem{ACGL}B.~Ananthanarayan, G.~Colangelo, J.~Gasser and H.~Leutwyler,\\
%``Roy equation analysis of pi pi scattering''
hep-ph/0005297.
For recent work on the mathematical properties of the Roy equations, see:\\
J.~Gasser and G.~Wanders,
%``One-channel Roy equations revisited,''
{\it Eur.\ Phys.\ J.}~C  {\bf 10} (1999) 159;\\
G.~Wanders,
%``The role of the input in Roy's equations for pi pi scattering,''
hep-ph/0005042.

\bibitem{dispersive methods}I.~Caprini,
%``Dispersive and chiral symmetry constraints on the light meson form
%factors,'' 
{\it Eur.\ Phys.\ J.}~C  {\bf 13} (2000) 471;\\
M.~Jamin, J.~A.~Oller and A.~Pich,
%``S-wave K pi scattering in chiral perturbation theory with resonances,''
hep-ph/0006045.

\bibitem{Gounaris Sakurai}
G.~J.~Gounaris and J.~J.~Sakurai,
%``Finite Width Corrections To The Vector Meson Dominance Prediction For Rho
%$\to$ E+ E-,'' 
{\it Phys.\ Rev.\ Lett.}~{\bf 21} (1968) 244.

\bibitem{Truong}
T.~N.~Truong, {\it Phys.~Rev.}~D {\bf 61} (1988) 2526;
%``Taylor's series and dispersion relation analyses of the vector pion  form
%factor and their comparison with perturbative and non perturbative 
%calculations,'' 
hep-ph/9809476;\\
%``Dispersion relation analyses of pion form-factor, chiral perturbation
%theory and unitarized calculations,'' 
hep-ph/0001271;
%``When is it possible to use perturbation technique in field theory?,''
hep-ph/0006302;\\
J.~A.~Oller, E.~Oset and J.~R.~Pelaez,
%``Non-perturbative approach to effective chiral Lagrangians and meson
%interactions,'' 
{\it Phys.\ Rev.\ Lett.}~{\bf 80} (1998) 3452;
%``Meson meson and meson baryon interactions in a chiral non-perturbative
%approach,'' 
{\it Phys.\ Rev.}~D  {\bf 59} (1999) 074001;\\
J.~A.~Oller, E.~Oset and A.~Ramos,\\
%``Chiral unitary approach to meson meson and meson baryon interactions  and
%nuclear applications,'' 
{\it Prog.\ Part.\ Nucl.\ Phys.}~{\bf 45} (2000) 157.

\bibitem{Wanders}
The consistency of the chiral representation for the 
$\pi\pi$ scattering amplitude with the constraints imposed by analyticity,
unitarity and crossing symmetry are discussed in:\\
B.~Ananthanarayan, D.~Toublan and G.~Wanders,\\
%``Consistency of the chiral pion-pion scattering amplitudes with axiomatic
%constraints,'' 
{\it Phys.\ Rev.}~D {\bf 51} (1995) 1093;
%``Low Energy Sum Rules For Pion-Pion Scattering and Threshold Parameters,''
{\it ibid.}~D {\bf 53} (1996) 2362;

\newpage
G.~Wanders,
%``Determination of the chiral pion pion scattering parameters: A  proposal,''
{\it Helv.\ Phys.\ Acta} {\bf 70} (1997) 287;\\
%``Chiral two-loop pion pion scattering parameters from crossing-symmetric
%constraints,'' 
{\it Phys.\ Rev.}~D  {\bf 56} (1997) 4328.

\bibitem{CGL}G.~Colangelo, J.~Gasser and H.~Leutwyler,
%``The $\pi \pi$ S-wave scattering lengths,''
hep-ph/0007112.

\bibitem{Nieves Amoros}
The numerical results for the scattering lengths quoted by Bijnens 
{\it et al.}$\,$\cite{BCEGS} are based on a matching at threshold, where the
chiral perturbation series 
converges less rapidly. The input used to determine the relevant
effective coupling constants is discussed in:\\
J.~Nieves and E.~Ruiz Arriola,
%``Error estimates for pi pi scattering threshold parameters in chiral
%perturbation theory to two loops,''
hep-ph/9906437;\\
G.~Amoros, J.~Bijnens and P.~Talavera,
%``Low energy constants from K(l4) form-factors,''
{\it Phys.~Lett.}~B {\bf 480} (2000) 71;
%``K(l4) form-factors and pi pi scattering,''
hep-ph/0003258.

\bibitem{Pislak}S.~Pislak {\it et al.}, ``A new measurement of $K^+\rightarrow
  \pi^+\pi^-e^+\nu$'', talk at  Laboratori Nazionali di
  Frascati, June 22, 2000.

\bibitem{DIRAC}
B.~Adeva {\it et al.}, Proposal to the SPSLC, CERN/SPSLC 95-1 (1995);\\
A.~Lanaro {\it et al.} [DIRAC collaboration], $\pi N$ {\it Newsletter} {\bf 15}
  (1999) 270. 

\bibitem{bound state}
J.~Gasser, V.~E.~Lyubovitskij and A.~Rusetsky,\\
%``Numerical analysis of the pi+ pi- atom lifetime in ChPT,''
{\it Phys.\ Lett.}~B {\bf 471} (1999) 244.

\bibitem{Stern Sazdjian Fuchs}
N.~H.~Fuchs, H.~Sazdjian and J.~Stern,
%``How to probe the scale of (anti-q q) in chiral perturbation theory,''
{\it Phys.\ Lett.}~B {\bf 269} (1991) 183;\\
J.~Stern, H.~Sazdjian and N.~H.~Fuchs,
%``What pi - pi scattering tells us about chiral perturbation theory,''
{\it Phys.\ Rev.}~D  {\bf 47} (1993) 3814;\\
M.~Knecht, B.~Moussallam, J.~Stern and N.~H.~Fuchs,\\
%``The Low-energy pi pi amplitude to one and two loops,''
{\it Nucl.\ Phys.}~B {\bf 457} (1995) 513;
%``Determination of Two-Loop $\pi\pi$ Scattering Amplitude Parameters,''
{\it ibid.}~B {\bf 471} (1996) 445.
\end{thebibliography}
\end{document}